\documentclass[runningheads,a4paper]{llncs}

\usepackage{latexsym}
\usepackage{setspace}
\usepackage{cancel}
\usepackage{listings}
\usepackage{graphicx}
\usepackage{appendix}
\usepackage{amssymb}
\usepackage{stmaryrd}
\usepackage{amsmath}
\usepackage{leftidx}
\usepackage{mathtools}
\usepackage[linesnumbered,noend]{algorithm2e}
\usepackage{paralist}

\usepackage{color}
\usepackage{wrapfig}
\usepackage{mathrsfs}
\usepackage{tikz}
\usetikzlibrary{shapes}
\usepackage{times}
\newcommand{\hide}[1]{}

\newcommand*\circled[1]{\overline{#1}}

\def\Cc{{\mathcal{C} }}

\def\Ee{{\mathcal{E} }}

\def\Ss{{\mathcal{S} }}

\def\schm{{\mathfrak{s} }}

\newcommand\wend{\rhd}

\newcommand\intnum{{\mathbb{Z} }}

\newcommand\intnumnz{\mathbb{Z}^{\neq 0}}

\newcommand\cur{\mathsf{cur}}
\newcommand\nnext{\mathsf{next}}
\newcommand\head{\mathsf{hd}}
\newcommand\tail{\mathsf{tl}}
\newcommand\init{\mathsf{init}}

\newcommand{\loopL}[1]{\mbox{loop\{} #1\mbox{\}}}
\newcommand{\ite}[3]{\mbox{if } (#1) \{#2\} \mbox{ [else } \{#3\}\mbox{]} }

\newcommand\vars{\mathsf{vars}}

\newcommand\dom{\mathsf{dom}}

\newcommand\rng{\mathsf{rng}}

\newcommand\ltrue{\mathsf{true}}

\newcommand\lfalse{\mathsf{false}}

\newcommand\maxv{\mathsf{max}}

\newcommand\addeq{+\!\!=}

\newcommand{\eval}[2]{\llbracket#1\rrbracket_{#2}}

\newcommand{\sval}{\Omega}
\newcommand{\sumf}{\Theta}
\newcommand{\initval}{\sval}

\newcommand{\vard}{\mathfrak{d}}

\newcommand{\csta}{\alpha}
\newcommand{\cstb}{\beta}

\newcommand{\cste}{\varepsilon}
\newcommand{\cstl}{\lambda}

\newcommand{\gmlasso}{\mathfrak{m}}

\newcommand{\abs}{\mathsf{Abs}}

\newcommand{\absset}{\mathscr{A}}

\newcommand{\interval}[1]{[#1]}

\newcommand{\yfc}[1]{\color{blue} {YF: #1 :YF} \color{black}}

\title{The~Commutativity~Problem~of~the~MapReduce Framework: A Transducer-based Approach\thanks{We found some inaccuracies in the conference version, which are fixed in this long version. We suggest the readers to refer to this version.}}
\titlerunning{Commutativity of MapReduce: A Transducer-based Approach}

\institute{Institute of Information Science, Academia Sinica \and State Key Laboratory of Computer Science,\\ Institute of Software, Chinese Academy of Sciences }

\author{Yu-Fang Chen\inst{1}, Lei Song\inst{2}, Zhilin Wu\inst{2}}

\begin{document}

\maketitle

\vspace{-4mm}

\begin{abstract}

MapReduce is a popular programming model for data parallel computation. 
In MapReduce, the \emph{reducer} produces an output from a list of inputs. Due to the scheduling policy of the platform, the inputs may arrive at the reducers in different order. The \emph{commutativity problem} of reducers asks if the output of a reducer is independent of the order of its inputs. Although the problem is undecidable in general,
the MapReduce programs in practice are usually used for data analytics and thus require very simple control flow. 
By exploiting the simplicity, we propose a programming language for reducers where the commutativity problem is decidable. The main idea of the reducer language is to separate the control and data flow of programs and disallow arithmetic operations in the control flow.
The decision procedure for the commutativity problem is obtained through a reduction to the equivalence problem of \emph{streaming numerical transducers} (SNTs), a novel automata model over infinite alphabets introduced in this paper. The design of SNTs is inspired by streaming transducers (Alur and Cerny, POPL 2011). Nevertheless, the two models are intrinsically different since the outputs of SNTs are integers while those of streaming transducers are data words. 
The decidability of the equivalence of SNTs is achieved with an involved combinatorial analysis of the evolvement of the values of the integer variables during the runs of SNTs.
\end{abstract}

\vspace{-8mm}

\section{Introduction}
MapReduce is a  popular framework for data parallel computation. It has been adopted in various cloud computing platforms including Hadoop~\cite{Hadoop} and Spark~\cite{Spark}. In a typical MapReduce program, a \emph{mapper} reads from data sources and outputs a list of key-value pairs. 
The scheduler of the MapReduce framework reorganizes the pairs $(k, v_1), (k,v_2)\ldots(k,v_n)$ with the same key $k$ to a pair $(k,l)$, where $l$ is a list of values $v_1,v_2,\ldots,v_n$, and sends $(k,l)$ to a \emph{reducer}. The reducer then iterates through the list and outputs a key-value pair~\footnote{We focus on the Hadoop style reducer in this work.}.
More specifically, taking the ``word-counting'' program as an example. It counts the occurrences of each word in a set of documents. The mappers read the documents and output for each document a list in the form of $(word_1, count_1)$, $(word_2, count_2)$, $\ldots$ , $(word_n, count_n)$, where $count_k$ is the number of occurrences of $word_k$ in the document being processed. These lists will be reorganized into the form of $(word_1, list_1), (word_2,list_2), \ldots, (word_n,list_n)$ and sent to the reducers, where $list_k$ is a list of integers recording the number of occurrences of $word_k$. Note that the \emph{order} of the integers in the lists can differ in different executions due to the scheduling policy. This results in the \emph{commutativity problem}.

A reducer is said to be \emph{commutative} if its output is independent of the order of its inputs. The commutativity problem asks if a reducer is commutative. A study from Microsoft~\cite{XZZ+14} reports that 58\% of the 507 reducers submitted to their MapReduce platform are non-commutative, which may lead to very tricky and hard-to-find bugs.
As an evidence, those reducers already went through serious code review, testing, and experiments with real data for months. Still, among them 5 reducers containing very subtle bugs caused by non-commutativity (confirmed by the programmers). 

The reducer commutativity problem in general is undecidable. However, in practice, MapReduce programs are usually used for data analytics and have very simple control structures. Many of them just iterate through the input list and compute the output with very simple operations. We want to study if the commutativity problem of real-world reducers is decidable. It has been shown in~\cite{CHSW15} that even with a simple programming language where the only loop structure allowed is to go over the input list once, the commutativity problem is already undecidable. Under scrutiny, we found that the language is still too expressive for typical data analytics programs. For example, it allows arbitrary multiplications of variables, which is a key element in the undecidability proof.


\smallskip

\noindent {\it Contributions}.
By observing the behavioral patterns of reducer programs for data analytics, we first design a programming language for reducers to characterize the essential features of them. 
We found that the commutativity problem becomes decidable if we partition variables into \emph{control variables} and \emph{data variables}. Control variables can occur in transition guards, but can only store values directly from the input list (e.g., it is not allowed to store the sum of two input values in a control variable). On the other hand, data variables are used to aggregate some information for outputs (e.g. sum of the values from the input list), but cannot be used in transition guards. This distinction is inspired by the streaming transducer model~\cite{RP11}, which, we believe, provides good insights for reducer programming language design in the MapReduce framework. Moreover, we assume that there are no nested loops in the language for reducers, which is a typical situation for MapReduce programs in practice.

We then introduce a formalism called \emph{streaming numerical transducers (SNT)} and obtain a decision procedure for the commutativity problem of the aforementioned language for reducers.
Similar to the language for reducers, SNTs distinguish between control variables and data variables. Although conceptually SNTs are similar to streaming transducers over data words introduced in \cite{RP11}, they are intrinsically different in the following sense: The outputs of SNTs are integers and the integer variables therein are manipulated by linear arithmetic operations. On the other hand, the outputs of streaming transducers are data words, and the data word variables are manipulated by concatenation operations. SNTs in this paper are assumed to be \emph{generalized flat}, which generalizes the ``flat'' automata (c.f. \cite{LS06}) in the sense that each nontrivial strongly connected component (SCC) of the transition graph is a collection of cycles, instead of one single cycle. Generalized flat transition graphs are sufficient to capture the transition structures of the programs in the aforementioned language for reducers.

The decision procedure for the commutativity problem is obtained by reducing to the equivalence problem of SNTs, which is further reduced to the non-zero output problem. The non-zero output problem asks whether given an SNT, there exists some input data word $w$ and initial valuation of variables such that the output of the SNT on $w$ is defined and non-zero.  For the non-zero output problem of SNTs, we apply a nontrivial combinatorial analysis of the evolvement of the integer variables during the runs of SNTs (Section~\ref{sec-sum}). The key idea of the decision procedure is that, generally speaking, if only the non-zero output problem is concerned, the different cycles in the SCCs can be dealt with \emph{independently} (Section~\ref{sec-glasso} and \ref{sec-gflat}). 
As a further evidence of the usefulness of SNTs for MapReduce programs, we demonstrate that SNTs can be composed to model and analyze the reducer programs that read the input list multiple times (Section~\ref{sec:cases}). 

As a novel formalism over infinite alphabets, the model of SNTs is interesting in its own right: On the one hand, SNTs are expressive in the sense that they include linear arithmetic operations on integer variables, while at the same time admit rather general transition graphs, that is, generalized flat transition graphs. On the other hand, despite this strong expressibility, it turns out that the commutativity problem, the equivalence problem, and the non-zero output problem of SNTs are still decidable.  

\smallskip

\noindent {\it Related work}.
SNTs can be seen as generalizations of register automata~\cite{KF94,NSV04} where registers correspond to the control variables in our terminology. Although register automata can have very general transition graphs beyond the generalized flat ones, they do not allow arithmetic operations on the variables.
There have been many automata models that contain arithmetic operations. Counter automata contain counters whose values can be updated by arithmetic operations (see \cite{Iba78,CJ98,LS06,HH14,FGH13}, to cite a few) in each transition.  Intuitively, the major difference between SNTs and counter automata is that SNTs work on data words and can apply arithmetic operations to an unbounded number of independent integer values, whereas  counter automata contain a bounded number of counters which involve only a bounded number of integer values in one configuration. Cost register automata (CRA)~\cite{ADD+13} also contain arithmetic operations, where the costs are stored into registers for which arithmetic operations can be applied. The equivalence of CRAs with the addition operation is decidable. SNTs are different from CRAs since the inputs of CRAs are words on finite alphabets, while those of SNTs are data words. 
Moreover, SNTs allow guards over variables ranging over an infinite domain but CRAs do not. 
%
%
There have been several transducer models on data words: Streaming transducers~\cite{RP11} mentioned before and symbolic transducers~\cite{VHL+12}. Symbolic transducers have data words as both inputs and outputs. They can put guards on the input value in one position of data words, but are incapable of comparing and aggregating multiple input values in different positions. 
In~\cite{NSS+15}, the authors considered a model for reducers in the MapReduce framework where
the only comparison that can be performed between data values are equalities, and
the reducers are essentially register automata/transducers.
Their model can describe a system with multiple layers of mappers and reducers.

The rest of the paper is organized as follows. Section~\ref{sec:preliminaries} defines the notations used in this paper. Section~\ref{sec:language} describes our design of the programming language for reducers. Section~\ref{sec:def-snt} defines SNTs. Section~\ref{sec:dec-snt} describes the decision procedure of SNTs. Section~\ref{sec:cases} discusses how to use our approach to analyze the commutativity property of more challenging data analytics programs. We conclude this work in Section~\ref{sec:conclusion}. The missing technical details and proofs can be found in the full version of this paper \cite{full-version}.

\vspace{-2mm}

\section{Preliminaries}
\label{sec:preliminaries}
Let $\intnum$,  $\intnumnz$ be the set of integers, non-zero integers, respectively.
We assume that all variables range over $\intnum$.  
For a function $f$, let $\dom(f)$ and $\rng(f)$ denote the \emph{domain}  and \emph{range} of $f$, respectively. 

An \emph{expression} $e$ over the set of variables $Z$ is defined by the following rules, $e::=  c \mid  c z \mid (e + e) \mid (e - e)$, where $z \in Z$ and $c\in \intnum$.  As a result of the commutativity and associativity of $+$, without loss of generality, we assume that all expressions $e$ in this paper are of the form $c_0 + c_1 z_1 + \dots + c_n z_n$, where $c_0, c_1,\dots,c_n \in \intnum$ and $z_1,\dots,z_n \in Z$. 
For an expression $e=c_0+c_1 z_1 + \dots + c_n z_n$, let $\vars(e)$ denote the set of variables $z_i$ such that $c_i \neq 0$. Let $\Ee_Z$ denote the set of all expressions over the set of variables $Z$. In this paper, it is assumed that all the constants in the expressions are encoded in binary.


A \emph{valuation} $\rho$ of $Z$ is a function from $Z$ to $\intnum$. A \emph{symbolic valuation} $\sval$ of $Z$ is a function that maps a variable in $Z$ to an expression (possibly over a different set of variables). The value of $e$ under a valuation $\rho$ (resp. symbolic valuation $\sval$), denoted by $\eval{e}{\rho}$ (resp. $\eval{e}{\sval}$), is defined recursively in the standard way. For example, let $\sval$ be a symbolic valuation the maps $z_1$ to $z_1+z_2$ and $z_2$ to $3z_2$, then $\eval{2z_1+z_2}{\sval}=2\eval{z_1}{\sval}+\eval{z_2}{\sval}=2(z_1+z_2)+3z_2=2z_1+5z_2$.
For a valuation $\rho$, a variable $z$, and $c \in \intnum$, define the valuation $\rho[c/z]$ such that $\rho[c/z](z)=c$ and $\rho[c/z](z')=\rho(z')$ for $z'\neq z$. 

In this paper, we use $X$ and $Y$ to denote the sets of \emph{control variables} and \emph{data variables}, respectively. We use the variable $\cur \notin X\cup Y$ to store the data value that is currently being processed in the input list and use $X^+$ to denote the set $X\cup \{\cur\}$.

A \emph{guard} over $Z$ is a formula defined by the rules $g::= \ltrue \mid x_1 = x_2 \mid x_1 > x_2 \mid x_1 < x_2 \mid g \wedge g$, where $x_1,x_2 \in Z$.
%
%
Let $\rho$ be a valuation of $X^+$ and $g$ be a guard over $X^+$ Then $\rho$ satisfies $g$, denoted by $\rho \models g$, iff $g$ is evaluated to $\ltrue$ under $\rho$. 
Let $\interval{n}$ denote the set $\{ 1, 2, \dots, n \}$, and $\interval{a,b}$ denote the set $\{ a, a+1, \dots, b \}$ when $b\geq a$ and $\emptyset$ otherwise. A \emph{permutation} on
$\interval{n}$ is a bijection from $\interval{n}$ to
$\interval{n}$. The set of
permutations on $\interval{n}$ is denoted by $S_n$.

A \emph{data word $w$} is a sequence of integer values $d_1\dots d_n$ such that $d_i \in \intnum$ for each $i$.
We use $\head(w)$, $\tail(w)$, and $|w|$ to denote the data value $d_1$, the tail $d_2\dots d_n$, and the length $n$, respectively.
We use $\epsilon$ to denote an empty data word. As a convention, we let $\head(\epsilon)=\bot$, $\tail(\epsilon)=\bot$, and $|\epsilon|=0$.
Given two data words $w,w'$, we use $w.w'$ to denote their concatenation.
Given $\sigma \in S_n$, we lift $\sigma$ to data words by defining $\sigma(w)=d_{\sigma(1)} \dots d_{\sigma(n)}$, for each data word $w=d_1\dots d_n$. We call $\sigma(w)$ as a permutation of $w$.

\vspace{-2mm}
 
\section{Language For Integer Reducers}\label{sec-mr-prog}
\label{sec:language}
We discuss the rationale behind the design of the programming language for reducers such that the commutativity problem is decidable. The language intends to support the following typical behavior pattern of reducers: A reducer program iterates through the input data word once, aggregates intermediate information into variables, and produces an output when it stops. 
%
Later in Section~\ref{sec:cases}, we will show an extension that allows resetting the iterators so that an input data word can be traversed multiple times.
\vspace{-4mm}
\begin{figure}
	\centering
	\begin{tabular}{rcl}
        $ s \in Statements$&$::=$&$y := e;\mid y \addeq e; \mid x:=x';\mid s\ s \mid \ite{g}{s}{s}$\\
		$ p\in Programs$&$::=$&$\loopL{s\ \nnext;}\mbox{ret }r; \mid s\ \nnext;p$		
	\end{tabular}
	\caption{A Simple Programming Language for Reducers. Here $x\in X$ are control variables, $y\in Y$ are data variables, $x' \in X^+$, $e\in \Ee_{X^+}$ are expressions, and $r$ is an expression in $\Ee_{X \cup Y}$. The square brackets mean that the else branch is optional. 
	}
	\label{fig:language}
\end{figure}

\vspace{-4mm}

More concretely, we focus on the programming language in Fig.~\ref{fig:language}. The language includes the usual features of program languages, variable assignments, sequential compositions, and conditional branchings. It also includes a statement $\nnext;$ which is used to advance the data word iterator. The $\loopL{s\ \nnext;}$ statement repeatedly executes the loop body $s\ \nnext;$ until reaching the end of input data word.
The novel feature of the language is that we partition the variables into two sets: \emph{control variables} $X$ and \emph{data variables} $Y$.
The variables from $X$ are used for guiding the control flow and the variables from $Y$ are used for storing aggregated intermediate data values.
The variables from $X$ can store only either initial values of variables in $X$ or values  occurring in the input data word. They can occur both in guards $g$ or arithmetic expressions $e$.
On the other hand, the variables from $Y$ can aggregate the results obtained from arithmetic expressions $e$, but cannot occur in guards $g$ or arithmetic expressions $e$. The initial values of variables can be arbitrary.
Given a program $p$, a data word $w$, and a valuation $\rho_0$, we use $p_{\rho_0}(w)$ to denote the output of $p$ on $w$, with the initial values of variables given by $\rho_0$. The formal semantics of the language can be found in the appendix.

Note that we do not allow multiplications in the language, so the reduction from the Diophantine equations in \cite{CHSW15} no longer works. Even though, if we do not distinguish the control and data variables, we can show easily that commutativity problem for this language is still undecidable, by a reduction from the reachability problem of Petri nets with inhibitor arcs~\cite{Min71,Rei08}.
The reachability problem of Petri nets with inhibitor arcs is reduced to the reachability problem of the reducer programs, which is in turn easily reduced to the commutativity problem of reducer programs.

Notice that in the programming language, we only allow additions ($\addeq$) or assignments ($:=$) of a new value computed from an expression over $X^+$ to data variables. 
In Fig.~\ref{fig:examples} we demonstrate a few examples performing data analytics operations. Observe that all of them follow the same behavioral pattern: The program iterates through the input data word and aggregates some intermediate information into some variables. The operations used for the aggregation are usually rather simple: either a new value is added to the variable (e.g. \texttt{sum} and \texttt{cnt} in Fig.~\ref{fig:examples}) storing the aggregated information, or a new value is assigned to the variable (e.g. \texttt{max} in Fig.~\ref{fig:examples}). Actually, the similar behavioral pattern occurs in all programs we have investigated.
Still, one may argue that allowing only additions and subtractions is too restrictive for data analytics. 
In Section~\ref{sec:cases}, we will discuss the extensions of the language to support more challenging examples, such as \emph{Mean Absolute Deviation} and \emph{Standard Deviation}.

\begin{figure}
	\centering
	\lstset{language=C,
		basicstyle=\ttfamily\scriptsize}
	\begin{tabular}{|c|c|}
		\hline
		\begin{minipage}[t]{0.4\textwidth}
		\vspace{-0.5cm}
			\begin{lstlisting}[mathescape=true]
max{
 max:=$\cur$;
 $\nnext$;
 loop{
  if ($\cur$>max)
   {max:=$\cur$;}
  $\nnext$;
 }
 ret max;}
	\end{lstlisting}
		\end{minipage}&
		\begin{minipage}[t]{0.4\textwidth}
		\vspace{-0.5cm}
			\begin{lstlisting}[mathescape=true]
sum{
 sum:=$\cur$;$\nnext$;
 loop{sum+=$\cur$;$\nnext$;}
 ret sum;}
			\end{lstlisting}
\hrule\vspace{0.1cm}%
			\begin{lstlisting}[mathescape=true]
cnt{
 cnt:=0;$\nnext$;
 loop{cnt+=1;$\nnext$;}
 ret cnt;}
			\end{lstlisting}			
		\end{minipage}\\
		\hline		
	\end{tabular}
	\caption{Examples of Reducers Performing Data Analytics Operations}
	\label{fig:examples}
	\vspace{-4mm}
\end{figure}


We focus on the following problems of reducer programs: (1) \emph{Commutativity}: given a program $p$, decide whether for each data word $w$ and its permutation $w'$, it holds that $p_{\rho_0}(w) = p_{\rho_0}(w')$ for all initial valuations $\rho_0$. (2) \emph{Equivalence:} given two programs $p$ and $p'$, decide whether for each data word $w$ and each initial valuation $\rho_0$, it holds that $p_{\rho_0}(w)=p'_{\rho_0}(w)$.

\vspace{-0.3cm}


\section{Streaming Numerical Transducers}\label{sec:def-snt}

In this section, we introduce \emph{streaming numerical transducers} (SNTs), whose inputs are data words and outputs are integer values. 
In SNT, we assume all data words end at a special symbol $\wend$, i.e., in the form of $\intnum^*\wend$.
A SNT scans a data word from left to right, records and aggregates information using control and data variables, and outputs an integer value when it finishes reading the data word. We will use SNTs to decide the commutativity and equivalence problem of the reducer programs defined in Section~\ref{sec-mr-prog}. 

A SNT $\Ss$ is a tuple $(Q, X, Y, \delta, q_0, O)$, where $Q$ is a finite set of states, $X$ is a finite set of control variables to store data values that have been met, $Y$ is a finite set of data variables to aggregate information for the output, $\delta$ is the set of transitions, $q_0 \in Q$ is the initial state, $O$ is the output function, which is a partial function from $Q$ to $\Ee_{X \cup Y}$.
The set of transitions $\delta$ comprises 
\begin{itemize}
\item the tuples $(q,  g \wedge \cur \neq \wend, \eta, q')$, where $q,q'\in Q$, $g$ is a guard over $X^+$ (defined in Section~\ref{sec:preliminaries}), and $\cur \neq \wend$ denotes the fact that the current position is not the end of the input,  and $\eta$ is an assignment function which is a partial function mapping $X \cup Y$  to $\Ee_{X^+ \cup Y}$ such that for each $x \in \dom(\eta) \cap X$, $\eta(x)=\cur$,
\item and the tuples $(q, g \wedge \cur = \wend, \eta, q')$, where $q,q' \in Q$, $g$ is a guard over $X$, and $\eta$ is an assignment function such that $\cur \not \in \rng(\eta)$ (they are called the $\triangleright$-transitions).
\end{itemize}
We write $q \xrightarrow{(g,\eta)} q'$ to denote $(q,g,\eta,q') \in \delta$ for convenience.  

In the following, we assume that for each $\triangleright$-transition $(q, g \wedge \cur = \wend, \eta, q')$, $q'$ is a sink-state, that is, that are no transitions out of $q'$. Moreover, we adopt the convention that when we mention the transition graph of an SNT $\Ss$, we always \emph{ignore} the $\triangleright$-transitions. 

In this paper, if not explicitly stated, we always assume that an SNT $\Ss$ satisfies the following additional constraints. (1) \emph{Deterministic:} For each pair of distinct transitions originating from $q$, say $(q, g_1, \eta_1,q'_1)$ and $(q, g_2,\eta_2,q'_2)$, it holds that $g_1 \wedge g_2$ is unsatisfiable. (2) \emph{Generalized flat:} Each SCC (strongly connected component) $S$ of the transition graph of $\Ss$ is either a single state or a set of simple cycles $\{C_1,\dots, C_n\}$ such that there is a state $q$ satisfying that each cycle $C_i$ (where $1 \le i \le n$) is of length one and is a self-loop around $q$,
%
(3) \emph{Independently evolving and copyless:} For each $(q, g, \eta, q') \in \delta$ and for each $y \in \dom(\eta) \cap Y$, $\eta(y)=e$ or $\eta(y)=y+e$ for some expression $e$ over $X^+$, (4) \emph{Monotone:}   For each control variable $x \in X$ and each nontrivial SCC $S$ of the transition graph,  when staying in $S$, either the value of $x$ is unchanged, or $x$ computes the maximum or minimum value. Formally, 
for each nontrivial SCC $S$ of the transition graph, the following conditions hold.
\begin{enumerate}
\item Each transition $(q, g, \eta, q)$ in $S$ satisfies that $g$ is a conjunction of the formulae of the form $\cur = x$, $\cur < x$, or $\cur > x$, where $x \in X$, and for each $x' \in \dom(\eta)$, $\eta(x')=\cur$.

\item For each control variable $x \in X$, the following constraints hold,
\begin{itemize}
\item for each self-loop $(q, g, \eta, q)$ in $S$, it holds that $\cur = x$, $\cur< x$, or $\cur > x$ is a conjunct of $g$, 
\item either all the self-loops $(q, g, \eta, q)$ such that $\cur > x$ is a conjunct of $g$ satisfy that $\eta(x)=\cur$, or none of them satisfies this constraint, similarly for the self-loops where $\cur < x$ occurs,

%
\item there do not exist self-loops $(q, g_1, \eta_1, q)$ and $(q, g_2, \eta_2,q)$ in $S$ such that $\cur > x$ is a conjunct of $g_1$, $\eta_1(x)=\cur$, $\cur < x$ is a conjunct of $g_2$, and $\eta_2(x)=\cur$.
%
\end{itemize}
\end{enumerate}

The semantics of an SNT $\Ss$  is defined as follows. A \emph{configuration} of $\Ss$ is a pair $(q,\rho)$, where $q \in Q$ and $\rho$ is a valuation of $X \cup Y$. An \emph{initial} configuration of $\Ss$ is $(q_0,\rho_0)$, where $\rho_0$ assigns arbitrary values to the variables from $X\cup Y$.
A sequence of configurations $(q_0,\rho_0)(q_1,\rho_1)\ldots(q_n,\rho_n)$ is
a \emph{run} of $\Ss$ over a data word $w=d_1 \dots d_n \wend$ iff there exists a path (sequence of transitions) $P=q_0 \xrightarrow{(g_1,\eta_1)} q_1 \xrightarrow{(g_2,\eta_2)} q_2 \dots q_{n-1} \xrightarrow{(g_n, \eta_n)} q_n \xrightarrow{(g_{n+1}, \eta_{n+1})} q_{n+1}$ such that for each $i \in [n+1]$, $\rho_{i-1}[d_i/\cur] \models g_i$, and $\rho_i$ is obtained from $\rho_{i-1}$ as follows: (1) For each $x \in X$, if $\eta_i(x)=\cur$ then $\rho_i(x)=d_i$,  otherwise $\rho_i(x)=\rho_{i-1}(x)$. (2) For each $y \in Y$, if $y \in \dom(\eta_i)$, then $\rho_i(y)=\eval{\eta_i(y)}{\rho_{i-1}[d_i/\cur]}$, otherwise, $\rho_i(y)=\rho_{i-1}(y)$.
We call $(q_{n+1},\rho_{n+1})$ the \emph{final configuration} of the run. In this case, we also say that the run follows the path $P$.
We say that a path $P$ in $\Ss$ is \emph{feasible} iff there exists a run of $\Ss$ following $P$. 


Given a data word $w = d_1 \dots d_n\wend$ and an initial configuration $(q_0, \rho_0)$, if there is a run of $\Ss$ over $w\wend$ starting from $(q_0,\rho_0)$ and with the final configuration $(q_{n+1},\rho_{n+1})$, then the output of $\Ss$ over $w\wend$ w.r.t. $\rho_0$, denoted by ${\Ss}_{\rho_0}(w\wend)$, is $\eval{O(q_{n+1})}{\rho_{n+1}}$. Otherwise, ${\Ss}_{\rho_0}(w\wend)$ is undefined, denoted by $\bot$.

\begin{example}[SNT for max]
The SNT $\Ss_{\max}$ for computing the maximum value of an input data word is defined as $(\{q_0,q_1,q_2\}, \{\maxv\}, \emptyset, \delta, q_0, O)$, where the set of transitions $\delta$ and the output function $O$ are illustrated in Fig.\ref{fig-snt-exmp}
%
(here $X=\{\maxv\}$, $Y=\emptyset$, and $\maxv:=\cur$ denotes the assignment of $\cur$ to the variable $\maxv$).

\vspace{-2mm}
\begin{figure}[htbp]
\begin{center}
\includegraphics{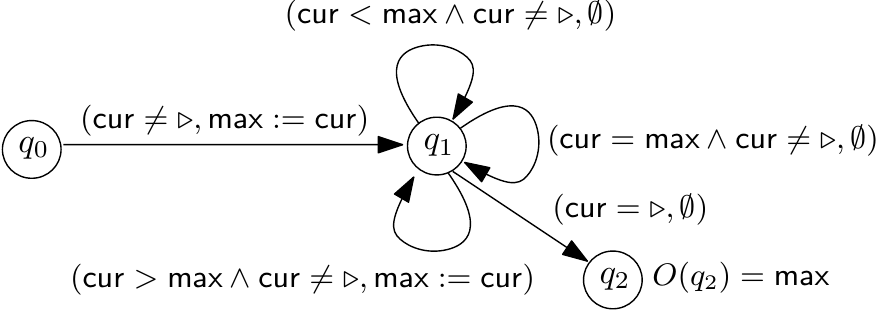}
\caption{The SNT $\Ss_{\max}$ for computing the maximum value}
\label{fig-snt-exmp}
\end{center}
\vspace{-6mm}
\end{figure}
\end{example}


\begin{proposition}\label{prop-mrprog-to-snt}
For each reducer program $p$, one can construct an equivalent SNT $\Ss_p$ which satisfies all the additional four constraints of SNTs, except the ``Monotone'' constraint.  Moreover, the number of states of $\Ss_p$ and the maximum number of simple cycles in an SCC of the transition graph of $\Ss_p$ are at most exponential in the number of branching statements in $p$. 
\end{proposition}
Intuitively, the main difference between reducer programs and SNTs are as follows: A reducer program moves to the next value of an input data word only when a $\nnext$ statement is executed, while an SNT advances the iterator in each transition.
As a result of this difference, when constructing $\Ss_p$ in Proposition~\ref{prop-mrprog-to-snt}, for each pair of consecutive ``$\nnext;$'' statements in $p$, the subprogram between them is transformed into a collection of transitions in $\Ss_p$, one for each execution path in the subprogram. Since the number of execution paths in a program is exponential in the number of branching statements therein, there is an exponential blow-up in the construction. For a reducer program $p$, the SNT $\Ss_p$ constructed in Proposition~\ref{prop-mrprog-to-snt} is not necessarily monotone. We define the class of \emph{monotone programs} as the class of reducer programs $p$ such that $\Ss_p$ is monotone. All the examples in Fig.~\ref{fig:examples} are monotone programs.

%
%
%

We focus on three decision problems of SNTs: (1) \emph{Commutativity}: Given an SNT $\Ss$, decide whether $\Ss$ is commutative, that is, whether for each data word $w\wend$ and each permutation $w'$ of $w$, $\Ss_{\rho_0}(w\wend)=\Ss_{\rho_0}(w'\wend)$ for all initial valuations $\rho_0$. (2) \emph{Equivalence}: Given two SNTs $\Ss,\Ss'$, decide whether $\Ss$ and $\Ss'$ are equivalent, that is, whether over each data word $w\wend$, $\Ss_{\rho_0}(w\wend)=\Ss'_{\rho_0}(w\wend)$ for all initial valuations $\rho_0$. (3) \emph{Non-zero output}: Given an SNT $\Ss$, decide whether $\Ss$ has a non-zero output, that is, whether there are a data word $w\wend$ and an initial valuation $\rho_0$ such that $\Ss_{\rho_0}(w\wend)\notin \{\bot, 0\}$. 

We first observe that the commutativity problem of SNTs can be reduced to the equivalence problem, which can be further reduced to the non-zero output problem of SNTs. For analyzing the complexity of the decision procedure in the next section,  we will state the complexity of the reductions w.r.t. the following factors of SNTs: the number of states, the number of control variables (resp. data variables), and the number of simple cycles of the transition graph. We will adopt the convention that if after a reduction, some factor becomes exponential, then this fact will be stated explicitly, and on the other hand, if some factor is still polynomial after the reduction, then this fact will be made implicit and will not be stated explicitly.


\vspace{-1mm}
\begin{proposition}\label{prop-snt-cmm-to-eqv}
The commutativity problem of SNTs is reduced to the equivalence problem of SNTs in polynomial time.
\end{proposition}

We briefly describe the idea of the reduction in Proposition~\ref{prop-snt-cmm-to-eqv} here. Suppose that $\Ss=(Q, X, Y, \delta, q_0, O)$ is an SNT such that $X=\{x_1,\dots,x_k\}$ and $Y=\{y_1,\dots,y_l\}$. Without loss of generality, we assume that the output of $\Ss$ is defined only for data words of length at least two. We will construct two SNTs $\Ss_1$ and $\Ss_2$ so that $\Ss$ is commutative iff $\Ss$ is equivalent to both $\Ss_1$ and $\Ss_2$.
\begin{itemize}
\item Intuitively, over a data word $w=d_1 d_2 d_3 \dots d_n\wend$ with $n\ge 2$, $\Ss_1$ simulates the run of $\Ss$ over $d_2 d_1 d_3 \dots d_n\wend$, that is, the data word obtained from $w\wend$ by swapping the first two data values. An additional write-once control variable $x'$ is introduced in $\Ss_1$ to store the data value $d_1$.
\item Intuitively, over a data word $w=d_1 d_2 d_3 \dots d_n\wend$ with $n\ge 2$, $\Ss_2$ simulates the run of $\Ss$ over $d_2 d_3 \dots d_n d_1\wend$, that is, the data word obtained from $w\wend$ by moving the first data value to the end. An additional write-once control variable $x'$ is introduced in $\Ss_2$ to store the data value $d_1$.
\end{itemize}
The correctness of this reduction follows from the fact that all the permutations of $d_1\dots d_n\wend$ can be generated by composing the two aforementioned permutations corresponding to $\Ss_1$ and $\Ss_2$ respectively (cf. Proposition 1 in \cite{CHSW15}). The construction of $\Ss_1$ (resp. $\Ss_2$) from $\Ss$ is in  polynomial time w.r.t. the size of $\Ss$.

Next, we reduce the equivalence problem of two SNTs $\Ss_1, \Ss_2$ to the non-zero output problem of another SNT $\Ss_3$. 

\begin{proposition}\label{prop-snt-eqv-to-nzero}
From two SNTs $\Ss_1$ and $\Ss_2$, an SNT $\Ss_3$ can be constructed in polynomial time such that  $(\Ss_1)_{\rho_0}(w\wend) \neq (\Ss_2)_{\rho_0}(w\wend)$ for some  data word $w \wend$ and valuation $\rho_0$  iff $(\Ss_3)_{\rho_0}(w\wend) \not\in \{\bot,0\}$ for some data word $w\wend$ and valuation $\rho_0$.
\end{proposition}
The SNT $\Ss_3$ can be constructed by the product construction. Note that the product construction preserves both the ``Generalized-flat'' and ``Monotone'' constraint. 

To facilitate the decision procedure in the next section, in the following, we will show for each SNT $\Ss$, the nonzero-output problem of $\Ss$ can be reduced to a series of nonzero-output problems of normalized SNTs $\Ss'$ which enjoy some nice properties. To this end, we introduce some additional notations below.

Let $P$ be a path in $\Ss$ from $q_0$ to some state $q$. We define ${\sf Conf}_P$ as the set of configurations $(q, \rho)$ such that there are a data word $w$ and an initial valuation $\rho_0$ satisfying that the run of $\Ss$ over $w$, starting from the initial configuration $(q_0, \rho_0)$, follows a path $P P'$ (i.e. $P$ is a prefix of the path $PP'$), and reaches the configuration $(q,\rho)$ after going through $P$. Let $x_1, x_2 \in X$. Then $P$ is said to \emph{enforce} $x_1 < x_2$ (resp. $x_1 = x_2, x_1 > x_2$) if for each $(q,\rho) \in {\sf Conf}_P$, it holds that $\rho \models x_1 < x_2$ (resp. $\rho \models x_1 = x_2$, $\rho \models x_1 > x_2$). 

An SNT $\Ss=(Q, X, Y, \delta, q_0, O)$ is said to be \emph{normalized} if (1) {\it Path-feasible:} for each path $P$ starting from $q_0$, ${\sf Conf}_P \neq \emptyset$, (2) {\it State-dominating:} all the paths starting from the initial state $q_0$ and ending at a given state $q$ (including the empty path $q_0$) enforce the same order relation between control variables, more precisely, for each state $q$ and each pair of distinct control variables $x_1, x_2 \in X$, all the paths  from $q_0$ to $q$ enforce $x_1 > x_2$, or all the paths from $q_0$ to $q$ enforce $x_1 < x_2$, or all the paths from $q_0$ to $q$ enforce $x_1 = x_2$, (3) {\it $\triangleright$-transition-guard-free}: Each $\triangleright$-transition $(q, g \wedge \cur = \triangleright, \eta, q')$ satisfies that $g = \ltrue$.

\begin{proposition}\label{prop-snt-norm}
For each SNT $\Ss$, the nonzero-output problem of $\Ss$ can be reduced to a series of the nonzero-output problems of normalized SNTs $\Ss'$ in exponential time.
\end{proposition}

The main idea of the reduction in Proposition~\ref{prop-snt-norm} is to record in the states the order relation between control variables, which is enforced by the path that has been traversed so far. Since a state in $\Ss$ may be split into several states in $\Ss'$, the transition graph of $\Ss'$ might not be generalized flat any more. Nevertheless, we can show that this will not happen and the ``Monotone  constraint'' guarantees that the normalized SNTs $\Ss'$ are still generalized flat. The SNT$\Ss_{\sf max}$  illustrated in Fig.~\ref{fig-snt-exmp} is normalized, since there is just one control variable in $\Ss_{\sf max}$.

In the rest of this paper, we assume that all the SNTs $\Ss=(Q, X, Y, \delta, q_0, O)$  are normalized. Moreover, for each $q \in Q$, we use $\preceq_q$ to denote the total preorder comprising the pairs $(x_i, x_j), (x_j, x_i) \in X \times X$ such that all the paths from $q_0$ to $q$ enforce $x_i = x_j$, the pairs $ (x_i, x_j)  \in X \times X$ such that all the paths from $q_0$ to $q$ enforce $x_i < x_j$. Let $\sim_q$ denote the equivalence relation on $[k]$ induced by $\preceq_q$, that is, $i \sim_q j$ iff $x_i \preceq_q x_j$ and $x_j \preceq_q x_i$. We assume that each normalized SNT $\Ss=(Q, X, Y, \delta, q_0, O)$ satisfies that for each $q \in Q$, $\preceq_q$ and $\sim_q$ can be computed from $q$ in linear time. In addition, we use $r^{\circled{q}}$ to denote the number of equivalence classes of $\sim_q$.

\vspace{-0.3cm}

\section{Decision procedure for the non-zero output problem}\label{sec:dec-snt}
We prove our main result, Theorem~\ref{thm:correctness}, by presenting a decision procedure for the non-zero output problem of normalized SNTs. We fix an SNT $\Ss = (Q,X,Y,\delta,q_0,O)$ such that $X=\{ x_1,\dots, x_k\}$ and $Y = \{y_1,\dots,y_l\}$. 
We first define summaries of the computations of $\Ss$ on paths and cycles in Section~\ref{sec-sum}, then present a decision procedure for the case that the transition graph of $\Ss$ is a \emph{generalized lasso} in Section~\ref{sec-glasso}. The transition graph of $\Ss$ is said to be a generalized lasso if it comprises a handle $H=q_0 \xrightarrow{(g_1,\eta_1)} q_1 \dots q_{m-1} \xrightarrow{(g_m,\eta_m)} q_{m}$, a collection of simple cycles $C_1,\dots,C_n$ such that each cycle $C_i$ is a self-loop around $q_m$, and a $\triangleright$-transition $q_m \xrightarrow{(\cur = \triangleright, \eta_{m+1})} q_{m+1}$. We extend the procedure to SNTs whose transition graphs are not necessarily generalized lassos in Section~\ref{sec-gflat}.

\begin{theorem}\label{thm:correctness}
The non-zero output problem of normalized SNTs can be decided in time exponential in the number of control and data variables and the number of simple cycles of the transition graph.
\end{theorem}

\begin{corollary}\label{cor:snt-dec-proc}
The commutativity problem of  monotone reducer programs can be decided in time exponential in the number of control and data variables, and doubly exponential in the number of branching statements of reducer programs. 
\end{corollary}

\begin{remark}
Though the decision procedure for the commutativity problem of monotone reducer programs has a complexity exponential in the number of data variables, and doubly exponential in the number of branching statements, we believe that the decision procedure could still be implemented to automatically analyze the programs in practice, in which these numbers are usually small. 
\end{remark}


\subsection{Summarization of the computations on paths and cycles}\label{sec-sum}

Suppose $P=p_0 \xrightarrow{(g_1,\eta_1)} p_1 \dots p_{n-1} \xrightarrow{(g_n,\eta_n)} p_{n}$ is a path of $\Ss$. We assume that the initial values of the control and data variables are represented by a symbolic valuation $\sval$ over $X \cup Y$ such that for each pair of variables $x_i, x_j \in X$, $\sval(x_j)=\sval(x_j)$ iff $x_i \sim_{p_0} x_j$. When $P$ is traversed in a run of $\Ss$ over a data word $w$,  the data value in a position of $w$ may have to be (un)equal to the initial value of some control variable or some other data value in $w$ that have been met before (enforced by the guards and assignments in $P$). Let $\sim_P$ denote the equivalence relation on $[n+k]$ induced by $P$ defined as follows: 
\begin{itemize}
\item For each $i, j \in [k]$, $i \sim_P j$ iff $x_i \sim_{p_0} x_j$.
\item For each $i, j \in [n]$, $k+i \sim_P k+j$ iff the guards and assignments on $P$ enforce that the data value in the $i$-th position of $w$ must be equal to that in the $j$-th position of $w$.
\item For each $i \in [k]$ and $j \in [n]$, $i \sim_P k+j$ iff the guards and assignments on $P$ enforce that the data value in the $j$-th position of $w$ must be equal to the initial value of $x_i$. 
\end{itemize}
Assuming that there are $r^{\circled{P}}$ ``\emph{fresh}'' equivalence classes of $\sim$, that is, equivalence classes $J$ of $\sim_P$ such that $J \cap [k] = \emptyset$ (intuitively, the data value represented by $J$ is not enforced to be equivalent to the initial values of control variables). 
We use the variables $\vard^{\circled{P}}_1,\vard^{\circled{P}}_2,\dots, \vard^{\circled{P}}_{r^{\circled{P}}}$ to denote the data values corresponding to these ``fresh'' equivalence classes, one for each such equivalence class. Note here we use the superscript ${\circled{P}}$ to denote the fact that $r^{\circled{P}}$ (resp. $\vard^{\circled{P}}_1$, $\dots$) is associated with the path $P$. In addition, we assume that there are $s^{\overline{p_0}}$ equivalence classes of $\sim_P$ on $[k]$, that is, equivalence classes $J$ of $\sim_P$ on $[n+k]$ such that $J \cap [k] \neq \emptyset$. Suppose $J_1,\dots, J_{s^{\circled{p_0}}}$ is an enumeration of these equivalence classes of $\sim_P$ on $[k]$. Let $\pi^{\circled{p_0}}: [s^{\circled{p_0}}] \rightarrow [k]$ such that $\pi^{\circled{p_0}}(j) = \min(J_j \cap [k])$ for each $j \in [s^{\circled{p_0}}]$. Intuitively, $\pi^{\circled{p_0}}$ chooses a representative control variable for each equivalence class. Note that $\pi^{\circled{p_0}}$ is an injective function, moreover, $s^{\circled{p_0}}$ and $\pi^{\circled{p_0}}$ are completely determined by $\sim_{p_0}$.

\begin{example}
Let $\Ss$ be an SNT where $X=\{x\}$, $Y=\{y\}$, and $P = p_0 \xrightarrow{(g_1,\eta_1)} p_1 \xrightarrow{(g_2,\eta_2)} p_2  \xrightarrow{(g_3,\eta_3)} p_3$ be a path of $\Ss$ such that $(g_1,\eta_1) = (\cur = x, y:= \cur)$, $(g_2,\eta_2)= (\ltrue, (x:=\cur, y:=y+\cur))$, $(g_3,\eta_3)= (\cur=x, y:=y+\cur)$. Then $k=1$, $n=3$. The guards and assignments enforce that the data value in position $1$ is equal to the initial value of $x$, which implies that $1 \sim_P 1+1$, i.e. $1 \sim_P 2$, in addition, the data value in position $2$ is equal to that in position $3$, which implies that $1+2 \sim_P 1+3$, i.e. $3 \sim_P 4$. Therefore, the equivalence relation $\sim_P$ has two equivalence classes, $\{1,2\}$ and $\{3,4\}$, of which $\{3,4\}$ is the fresh equivalence class. We conclude that $r^{\circled{P}}=1$ and $\vard^{\circled{P}}_1$ is used to denote the data value corresponding to this fresh equivalence class.
\end{example}

\begin{proposition}\label{prop-sum-path}
Suppose that $P$ is a path starting form $p_0$ and the initial values of $X \cup Y$ are represented by a symbolic valuation $\initval$ such that for each pair of variables $x_i, x_j \in X$, $\sval(x_j)=\sval(x_j)$ iff $x_i \sim_{p_0} x_j$. Then the values of $X \cup Y$ after traversing the path $P$ are specified by a symbolic valuation $\sumf^{(P,\initval)}$ satisfying the following conditions.
\begin{itemize}
\item The set of indices of $X$, i.e., $[k]$, is partitioned into $I^{\circled{P}}_{pe}$ and $I^{\circled{P}}_{tr}$, the indices of \emph{persistent} and \emph{transient} control variables, respectively. A control variable is persistent if it stores the initial value of some control variable after traversing $P$, otherwise, it is transient.
\item For each $x_j\in X$ such that $j \in I^{\circled{P}}_{pe}$, $\sumf^{(P,\initval)}(x_j)=\sval(x_{\pi^{\circled{p_0}}(\pi^{\circled{P}}_{pe}(j))})$, where $\pi^{\circled{P}}_{pe}: I^{\circled{P}}_{pe} \rightarrow [s^{\circled{p_0}}]$ is a mapping from the index of a persistent control variable $x_j$ to the index of the equivalence class such that the initial value of control variables corresponding to this equivalence class is assigned to $x_j$ after traversing $P$.
\item  For each $x_j\in X$ such that $j\in I^{\circled{P}}_{tr}$,
$\sumf^{(P,\initval)}(x_j)=\vard^{\circled{P}}_{\pi^{\circled{P}}_{tr}(j)}$, where $\pi^{\circled{P}}_{tr}: I^{\circled{P}}_{tr} \rightarrow [r^{\circled{P}}]$ is a mapping from the index of a transient control variable to the index of the data value assigned to it.
\item For each $y_j \in Y$, 
\[
 \sumf^{(P,\initval)}(y_j)  =
 \cste^{\circled{P}}_{j} + 
 \cstl^{\circled{P}}_j \initval(y_j)  + 
  \sum\limits_{j'\in [s^{\circled{p_0}}]} \csta^{\circled{P}}_{j,j'}\initval(x_{\pi^{\circled{p_0}}(j')}) +
  \sum\limits_{j''\in [r^{\circled{P}}]}\cstb^{\circled{P}}_{j,j''} \vard^{\circled{P}}_{j''},
\]  
\hide{
\item For each $y_j \in Y$, 
\[
\small
\begin{array}{l}
\smallskip
\sumf^{(P,\initval)}(y_j)  = \\
\hspace{2mm} \cste^{\circled{P}}_{j} + \cstl^{\circled{P}}_j \initval(y_j)  + \csta^{\circled{P}}_{j,1} \initval(x_1) + \dots + \csta^{\circled{P}}_{j,k} \initval(x_k) +  \cstb^{\circled{P}}_{j,1} \vard^{\circled{P}}_1 +\dots + \cstb^{\circled{P}}_{j,r^{\circled{P}}} \vard^{\circled{P}}_{r^{\circled{P}}},
\end{array}
\]} 
where $\cste^{\circled{P}}_j,\cstl^{\circled{P}}_j, \csta^{\circled{P}}_{j,1},\dots,\csta^{\circled{P}}_{j, s^{\circled{p_0}}}, \cstb^{\circled{P}}_{j,1},\dots,\cstb^{\circled{P}}_{j,r^{\circled{P}}}$ are integer constants such that $\cstl^{\circled{P}}_{j} \in \{0,1\}$ (as a result of the ``independently evolving and copyless'' constraint).  It can happen that $\cstl^{\circled{P}}_j =0$,  which means that $\initval(y_j)$ is irrelevant to $\sumf^{(P,\initval)}(y_j)$. Similarly for $\csta^{\circled{P}}_{j,1}=0$, and so on.
\end{itemize}
\end{proposition}
In Proposition~\ref{prop-sum-path}, the sets $I^{\circled{P}}_{pe}$ and $I^{\circled{P}}_{tr}$, the mapping $\pi^{\circled{P}}_{pe}$ and $\pi^{\circled{P}}_{tr}$, and the constants $\cste^{\circled{P}}_j,\cstl^{\circled{P}}_j, \dots, \cstb^{\circled{P}}_{j,r^{\circled{P}}}$ only depend on $P$ and are independent of $\initval$. In addition, they can be computed in polynomial time from (the transitions in) $P$.
We define $(\pi^{\circled{P}}_{pe})^{-1}$ as the inverse function of $\pi^{\circled{P}}_{pe}$, that is, for each $j' \in \rng(\pi^{\circled{P}}_{pe})$, $(\pi^{\circled{P}}_{pe})^{-1}(j')=\{j \in I^{\circled{P}}_{pe}  \mid \pi^{\circled{P}}(j)= j'\}$. Similarly for $(\pi^{\circled{P}}_{tr})^{-1}$.

As a corollary of Proposition~\ref{prop-sum-path}, the following result demonstrates how to summarize the computations of $\Ss$ on the composition of two paths.

\begin{corollary}\label{cor-comp-two-paths}
Suppose that $P_1$ and $P_2$ are two paths in $\Ss$ such that the last state of $P_1$ is the same as the first state of $P_2$. Moreover, let $\sumf^{(P_1, \initval)}$ (resp. $\sumf^{(P_2, \initval)}$) be the symbolic valuation summarizing the computation of $\Ss$ on $P_1$ (resp. $P_2$). Then the symbolic valuation summarizing the computation of $\Ss$ on $P_1 P_2$ is $\sumf^{(P_2,\ \sumf^{(P_1,\initval)})}$.
\end{corollary}

Let the first state of $P_1$ and $P_2$ be $p_{1,0}$ and $p_{2,0}$ respectively. In order to get a better understanding of the relation between $\sumf^{(P_2,\ \sumf^{(P_1,\initval)})}$ and $(\sumf^{(P_1, \initval)},\sumf^{(P_2, \initval)})$, in the following, for each $y_j \in Y$, we obtain a more explicit form of the expression $\sumf^{(P_2,\ \sumf^{(P_1,\initval)})}(y_j)$, by unfolding therein the expression $\sumf^{(P_1,\initval)}$,\\
\medskip
\resizebox{\hsize}{!}{
	$\begin{array}{rl}
	\medskip
	\sumf^{(P_2,\ \sumf^{(P_1,\initval)})}(y_j) = & 
	\left(\cste^{\circled{P_2}}_{j}+
	\cstl^{\circled{P_2}}_{j} \cste^{\circled{P_1}}_{j}\right)+ \left(\cstl^{\circled{P_2}}_{j} \cstl^{\circled{P_1}}_{j} \right) \initval(y_j)\ +\\
	\medskip
	& \sum \limits_{j' \in \rng(\pi^{\circled{P_1}}_{pe})} 
	\left(\cstl^{\circled{P_2}}_{j} \csta^{\circled{P_1}}_{j,j'} + \sum \limits_{j'' \in (\pi^{\circled{P_1}}_{pe})^{-1}(j')\ \cap\ \rng(\pi^{\circled{p_{2,0}}})}  \csta^{\circled{P_2}}_{j,(\pi^{\circled{p_{2,0}}})^{-1}(j'')} \right) \initval(x_{\pi^{\circled{p_{1,0}}}(j')})\ + \\
	\medskip
	& 
	\sum \limits_{j' \in  [s^{\circled{p_{1,0}}}] \setminus \rng(\pi^{\circled{P_1}}_{pe}) } 
	\left(\cstl^{\circled{P_2}}_{j} \csta^{\circled{P_1}}_{j,j'} \right) \initval(x_{\pi^{\circled{p_{1,0}}}(j')})\ +\\
	\medskip
	& \sum \limits_{j' \in \rng(\pi^{\circled{P_1}}_{tr})} \left(\cstl^{\circled{P_2}}_{j} \cstb^{\circled{P_1}}_{j,j'} + \sum \limits_{j'' \in (\pi^{\circled{P_1}}_{tr})^{-1}(j') \cap \rng(\pi^{\circled{p_{2,0}}})} \csta^{\circled{P_2}}_{j, (\pi^{\circled{p_{2,0}}})^{-1}(j'')} \right) \vard^{\circled{P_1}}_{j'}\ +
	 \\
	& 
	\sum \limits_{j' \in [r^{\circled{P_1}}]\setminus \rng(\pi^{\circled{P_1}}_{tr})} \left( \cstl^{\circled{P_2}}_{j} \cstb^{\circled{P_1}}_{j,j'} \right) \vard^{\circled{P_1}}_{j'} +
	
	\sum \limits_{j'\in[r^{\circled{P_2}}]} \cstb^{\circled{P_2}}_{j,j'} \vard^{\circled{P_2}}_{j'}.
	\end{array}$
}\medskip\\
In the equation, $j' \in  \rng(\pi^{\circled{P_1}}_{pe})$ implies that for each $j'' \in  (\pi^{\circled{P_1}}_{pe})^{-1}(j')$,  $x_{j''}$ stores the initial value of $x_{\pi^{\circled{p_{1,0}}}(j')}$ after traversing $P_1$, which means that the initial value of $x_{j''}$ for each $j'' \in  (\pi^{\circled{P_1}}_{pe})^{-1}(j')$ before traversing $P_2$ is $\initval(x_{\pi^{\circled{p_{1,0}}}(j')})$, and some of $x_{j''}$ for $j'' \in  (\pi^{\circled{P_1}}_{pe})^{-1}(j')$ are chosen as the representatives of the equivalence classes of $\sim_{p_{2,0}}$ (in this case, $j''$ is in the range of $\pi^{\circled{p_{2,0}}}$), therefore we have the item $\left( \sum \limits_{j'' \in (\pi^{\circled{P_1}}_{pe})^{-1}(j')\ \cap\ \rng(\pi^{\circled{p_{2,0}}})}  \csta^{\circled{P_2}}_{j,(\pi^{\circled{p_{2,0}}})^{-1}(j'')} \right) \initval(x_{\pi^{\circled{p_{1,0}}}(j')})$. When $j' \in \rng(\pi^{\circled{P_1}}_{tr})$, the initial value of $x_{j''}$ for each $j'' \in (\pi^{\circled{P_1}}_{tr})^{-1}(j')$ before traversing $P_2$ is $\vard^{\circled{P_1}}_{j'}$, and some of $x_{j''}$ for $j'' \in (\pi^{\circled{P_1}}_{tr})^{-1}(j')$ are chosen as the representatives of equivalence classes of $\sim_{p_{2,0}}$, therefore we have the item $\left(\sum \limits_{j'' \in (\pi^{\circled{P_1}}_{tr})^{-1}(j') \cap \rng(\pi^{\circled{p_{2,0}}})} \csta^{\circled{P_2}}_{j, (\pi^{\circled{p_{2,0}}})^{-1}(j'')} \right)\ \vard^{\circled{P_1}}_{j'}$.
For $j'\in [s^{\circled{p_{1,0}}}] =\rng(\pi^{\circled{P_1}}_{pe}) \cup ([s^{\circled{p_{1,0}}}] \setminus \rng(\pi^{\circled{P_1}}_{pe}))$, we have the item $(\cstl^{\circled{P_2}}_{j} \csta^{\circled{P_1}}_{j,j'}) \initval(x_{\pi^{\circled{p_{1,0}}}(j')})$, i.e. the coefficient of $\initval(x_{\pi^{\circled{p_{1,0}}}(j')})$ in $\sumf^{(P_1, \initval)}$ multiplied by $\cstl^{\circled{P_2}}_{j}$. Moreover, for $j'\in [r^{\circled{P_1}}] = \rng(\pi^{\circled{P_1}}_{tr}) \cup ([r^{\circled{P_1}}] \setminus \rng(\pi^{\circled{P_1}}_{tr}))$, we have 
the item $( \cstl^{\circled{P_2}}_{j} \cstb^{\circled{P_1}}_{j,j'}) \vard^{\circled{P_1}}_{j'}$, i.e. the coefficient of $\vard^{\circled{P_1}}_{j'}$ in $\sumf^{(P_1, \initval)}$ multiplied by $\cstl^{\circled{P_2}}_{j}$.

In the following, by utilizing Proposition~\ref{prop-sum-path} and Corollary~\ref{cor-comp-two-paths}, for each path $C^{\ell}$ ($\ell \ge 1$) which is obtained by iterating a simple cycle $C = (q, g, \eta, q)$ for $\ell$ times, we illustrate how $\sumf^{(C^\ell,\initval)}$ is related to $\sumf^{(C, \initval)}$ and $\ell$. For convenience, we call $\ell$ a \emph{cycle counter variable}. It is easy to observe that both $I^{\circled{C}}_{pe}$ and $I^{\circled{C}}_{tr}$ are the union of the equivalence classes of $\sim_{q}$. From the ``Monotone'' constraint, we know that for each $x \in \dom(\eta)$, it holds that $\eta(x)=\cur$. Therefore, if $j \in I^{\circled{C}}_{pe}$, then $x_j$ still stores the initial value of $x_j$ after traversing $C$. 
This implies that for each $j' \in \rng(\pi^{\circled{C}}_{pe})$, let $\pi^{\circled{q}}(j')=j$, then $\pi^{\circled{C}}_{pe}(j)=j'$.  Therefore, for each $j' \in \rng(\pi^{\circled{C}}_{pe})$, the value of $x_{\pi^{\circled{q}}(j')}$ is unchanged after traversing $C$.

\begin{proposition}\label{prop-sum-cycle}
Suppose that $C$ is a simple cycle (i.e. a self-loop around a state $q$) and $P=C^{\ell}$ such that $\ell \ge 2$. Then the symbolic valuation $\sumf^{(C^\ell,\initval)}$ to summarize the computation of $\Ss$ on $P$ is as follows:  

\noindent
\medskip
\resizebox{0.95\hsize}{!}{
$
\begin{array}{l c l}
\sumf^{(C^\ell,\initval)}(y_j)  &= & 
\left(1 + \cstl^{\circled{C}}_{j} + \dots +(\cstl^{\circled{C}}_{j})^{\ell - 1} \right)\cste^{\circled{C}}_{j} + (\cstl^{\circled{C}}_{j})^\ell \initval(y_j)\ + \\
\smallskip
&& \sum \limits_{j' \in \rng(\pi^{\circled{C}}_{pe}) } \left(1+\cstl^{\circled{C}}_{j} + \dots +(\cstl^{\circled{C}}_{j})^{\ell - 1} \right) \csta^{\circled{C}}_{j,j'}\initval(x_{\pi^{\circled{q}}(j')}) \ + \\
& &  \sum \limits_{j' \in [s^{\circled{q}}] \setminus \rng(\pi^{\circled{C}}_{pe}) }  (\cstl^{\circled{C}}_{j})^{\ell - 1} \csta^{\circled{C}}_{j,j'} \initval(x_{\pi^{\circled{q}}(j')})\ +  \\
\smallskip
&&  \sum \limits_{j' \in \rng(\pi^{\circled{C}}_{tr})} \sum \limits_{s\in[\ell -1]}
\left(\cstl^{\circled{C}}_{j}\cstb^{\circled{C}}_{j,j'}+ \sum \limits_{j'' \in (\pi^{\circled{C}}_{tr})^{-1}(j') \cap \rng(\pi^{\circled{q}})} \csta^{\circled{C}}_{j, (\pi^{\circled{q}})^{-1}(j'')}  \right)
(\cstl^{\circled{C}}_{j})^{\ell-s-1}
\vard^{\circled{C , s}}_{j'}\ +\\
\smallskip
&& \sum \limits_{j' \in [r^{\circled{C}}] \setminus \rng(\pi^{\circled{C}}_{tr})}\sum \limits_{s\in[\ell -1]} \left((\cstl^{\circled{C}}_{j})^{\ell - s} \cstb^{\circled{C}}_{j,j'} \right) \vard^{\circled{C , s}}_{j'} + 
\sum \limits_{j' \in [r^{\circled{C}}] }  
 \cstb^{\circled{C}}_{j, j'} \vard^{\circled{C , \ell}}_{j'},
\end{array} 
$
}
\medskip\\
where the variables $\vard^{\circled{C , s}}_{1}$ for $s\in [\ell]$
 represent the data values introduced when traversing $C$ for the $s$-th time. 
\end{proposition}

From Proposition~\ref{prop-sum-cycle} and the fact that $\cstl^{\circled{C}}_{j} \in \{0, 1\}$, we have the following observation.
\begin{itemize}
\item If $\cstl^{\circled{C}}_{j}=0$, then
%
\medskip\\
\resizebox{0.9\hsize}{!}{$
\begin{array}{l c l}
\sumf^{(C^\ell,\initval)}(y_j)  & = & \cste^{\circled{C}}_{j} +  \sum \limits_{j' \in \rng(\pi^{\circled{C}}_{pe})} \csta^{\circled{C}}_{j,j'} \initval(x_{\pi^{\circled{q}}(j')})\ +
\\
&& \sum \limits_{j'  \in \rng(\pi^{\circled{C}}_{tr})} \left(\sum \limits_{j'' \in (\pi^{\circled{C}}_{tr})^{-1}(j') \cap \rng(\pi^{\circled{q}})} \csta^{\circled{C}}_{j, (\pi^{\circled{q}})^{-1}(j'')}  \right) \vard^{\circled{C , \ell  -  1}}_{j'} +  
\sum \limits_{j' \in [r^{\circled{C}}] }  
 \cstb^{\circled{C}}_{j, j'} \vard^{\circled{C , \ell}}_{j'}.
 \end{array}
 $
}
\hide{
\item If $r^{\circled{C}}=0$ and $\cstl^{\circled{C}}_{j}=0$, then 
\[
\sumf^{(C^\ell,\initval)}(y_j)   =  \cste^{\circled{C}}_{j} +  \sum \limits_{j' \in I^{\circled{C}}_{pe}} \csta^{\circled{C}}_{j,j'} \initval(x_{j'}).
\]
}
\item If $\cstl^{\circled{C}}_{j}=1$, then
\medskip\\
\resizebox{0.95\hsize}{!}{$
\begin{array}{l c l}
\sumf^{(C^\ell,\initval)}(y_j) & = &   \ell \cste^{\circled{C}}_{j}  + \initval(y_j) +   \sum  \limits_{j' \in \rng(\pi^{\circled{C}}_{pe}) } \ell \csta^{\circled{C}}_{j,j'}  \initval(x_{\pi^{\circled{q}}(j')}) + 
\sum \limits_{j' \in [s^{\circled{q}}] \setminus \rng(\pi^{\circled{C}}_{pe})} \csta^{\circled{C}}_{j,j'} \initval(x_{\pi^{\circled{q}}(j')})\ + \\
\smallskip
& & \sum \limits_{j' \in \rng(\pi^{\circled{C}}_{tr})} \sum \limits_{s\in[\ell -1]}
\left(\cstb^{\circled{C}}_{j,j'} + \sum \limits_{j'' \in (\pi^{\circled{C}}_{tr})^{-1}(j') \cap \rng(\pi^{\circled{q}})} \csta^{\circled{C}}_{j, (\pi^{\circled{q}})^{-1}(j'')}  \right) \vard^{\circled{C , s}}_{j'}\ +\\
& &  \sum \limits_{j' \in [r^{\circled{C}}]  \setminus \rng(\pi^{\circled{C}}_{tr}) }\sum \limits_{s\in[\ell -1]} 
\cstb^{\circled{C}}_{j,j'} \vard^{\circled{C , s}}_{j'} + \sum \limits_{j' \in [r^{\circled{C}}] }  
\cstb^{\circled{C}}_{j, j'} \vard^{\circled{C , \ell}}_{j'}.
\end{array}
$
}
\hide{
\item If $r^{\circled{C}}=0$ and $\cstl^{\circled{C}}_{j}=1$, then
\[
\sumf^{(C^\ell,\initval)}(y_j)  =    \ell \cste^{\circled{C}}_{j}  + \initval(y_j) +   \sum  \limits_{j' \in I^{\circled{C}}_{pe}} \ell \csta^{\circled{C}}_{j,j'}  \initval(x_{j'}).
\]
}
\end{itemize}

%
%
%
%

\begin{example}\label{exmp-sum}
Let $\Ss'_{\sf max}$ be the SNT illustrated in Fig.~\ref{fig-dec-proc-snt-exmp}, which is obtained from $\Ss_{\sf max}$ by replacing  the control variable $\sf max$ with  $x_1$ and introducing the data variables $y_1,y_2, y_3$. Consider the cycle $C_1$ in $\Ss'_{\sf max}$. Then $O(q_2)= a_0 + a_1 x_1 + b_1 y_1 + b_2 y_2 + b_3 y_3 = y_1 - 2y_2 + y_3$. Moreover, $I^{\circled{C_1}}_{pe} = \{1\}$, $I^{\circled{C_1}}_{tr} = \emptyset$, $\pi^{\circled{C_1}}_{pe}(1) = 1$, $\cstl^{\circled{C_1}}_1 = \cstl^{\circled{C_1}}_2 = 1$, and $\cstl^{\circled{C_1}}_3 = 0$. On the other hand, $I^{\circled{C_2}}_{pe} = \emptyset$, $I^{\circled{C_2}}_{tr} = \{1\}$, $\pi^{\circled{C_2}}_{tr}(1) = 1$, and $\cstl^{\circled{C_2}}_1 = \cstl^{\circled{C_2}}_2 = \cstl^{\circled{C_2}}_3  = 1$. Suppose $\ell \ge 2$. Let $ \vard^{\circled{C_1, 1}}_1, \dots, \vard^{\circled{C_1, \ell}}_1$ represent the data values introduced when traversing the path $C^\ell_1$, then
\[
\begin{array}{l c l}
\sumf^{(C^\ell_1, \initval)}(y_1) &= & \initval(y_1) + (4 \ell) \initval(x_1) +\ \ \vard^{\circled{C_1, 1}}_1 + \dots +\ \ \vard^{\circled{C_1, \ell}}_1,\\
\sumf^{(C^\ell_1, \initval)}(y_2) &=& \initval(y_2) + (2 \ell) \initval(x_1) + 2 \vard^{\circled{C_1, 1}}_1 + \dots + 2 \vard^{\circled{C_1, \ell}}_1,\\
\sumf^{(C^\ell_1, \initval)}(y_3) &=& \hspace{5.5cm} 3 \vard^{\circled{C_1, \ell}}_1.
\end{array}
\] 
On the other hand, let $ \vard^{\circled{C_2, 1}}_1, \dots, \vard^{\circled{C_2, \ell}}_1$ represent the data values introduced when traversing the path $C^\ell_2$, then 
\[
\begin{array}{l c l}
\sumf^{(C^\ell_2, \initval)}(y_1) &= & \initval(y_1) +\ \  \initval(x_1) +4 \vard^{\circled{C_2, 1}}_1 + \dots +4 \vard^{\circled{C_2, \ell-1}}_1 + 3 \vard^{\circled{C_2, \ell}}_1,\\
\sumf^{(C^\ell_2, \initval)}(y_2) &=& \initval(y_2) + 3 \initval(x_1) + 5 \vard^{\circled{C_2, 1}}_1 + \dots + 5 \vard^{\circled{C_2, \ell-1}}_1 + 2 \vard^{\circled{C_2, \ell}}_1,\\
\sumf^{(C^\ell_2, \initval)}(y_3) &=& \initval(y_3) + 5 \initval(x_1) + 6 \vard^{\circled{C_2, 1}}_1 + \dots + 6 \vard^{\circled{C_2, \ell-1}}_1 + \ \ \vard^{\circled{C_2, \ell}}_1.
\end{array}
\] 

\vspace{-2mm}
\begin{figure}[htbp]
\begin{center}
\includegraphics[scale=0.9]{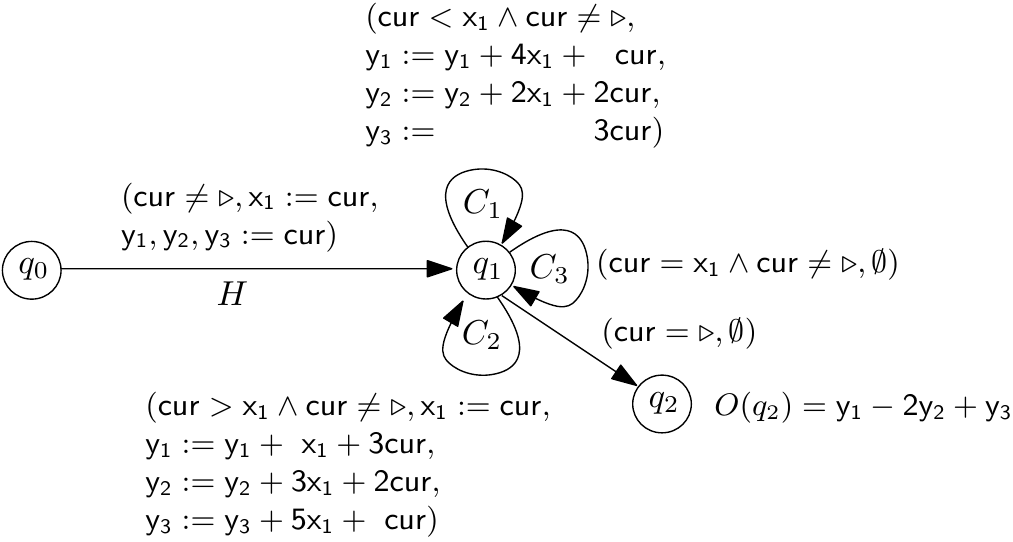}
\caption{The SNT $\Ss'_{\max}$: Extending $\Ss_{\max}$ with data variables}\label{fig-dec-proc-snt-exmp}
\end{center}
\end{figure}
\end{example}

\vspace{-2mm}

\subsection{Decision procedure for generalized lassos}\label{sec-glasso}

%
In this section, we present a decision procedure for SNTs whose transition graphs are generalized lassos. From Proposition~\ref{prop-sum-cycle}, we know that the coefficients containing the cycle counter variable $\ell$ in $\sumf^{(C^\ell,\initval)}(y_j)$ can be non-zero when $\cstl^{\circled{C}}_{j}=1$. The non-zero coefficients may propagate to the output expression.  In such a case, 
because the SNTs are ``transition-enabled'' (i.e. for any sequence of transitions, a corresponding run exists), %
intuitively, one can pick a run corresponding to a very large $\ell$ so that it dominates the value of the output expression and makes the output non-zero. 
In the decision procedure we are going to present, we first check if the handle of the generalized lasso produces a non-zero output in Step I.
We then check in Step II the coefficients containing $\ell$ in the output expression is non-zero. If this does not happen, then we show in Step III that the non-zero ouput problem of SNTs can be reduced to a finite state reachability problem and thus can be easily decided.

Before presenting the decision procedure, we introduce some notations.
Let $e$ be an expression consisting of symbolic values $\initval(z)$ for $z\in X\cup Y$ and variables $\vard_1, \dots, \vard_{s}$ corresponding to the values of the input data word. More specifically, let $e:=\mu_0 + \mu_1 \initval(z_1) +\dots + \mu_{k+l} \initval(z_{k+l}) + \xi_1 \vard_1 + \dots + \xi_{s} \vard_{s}$,
such that $\mu_0,\mu_1,\dots,\mu_{k+l}, \xi_1,\dots,\xi_{s}$ are expressions containing only constants and cycle counter variables.
Then we call $\mu_0$ the \emph{constant atom}, $\mu_i \initval(z_i)$ the $\initval(z_i)$-atom for $i\in[k+l]$, and $\xi_j \vard_j$ the $\vard_j$-atom for $j\in[s]$ of the expression $e$. Moreover, $\mu_1, \dots, \mu_{k+l}, \xi_1,\dots, \xi_{s}$ are called the \emph{coefficients} and $\initval(z_1), \dots, \initval(z_{k+l}), \vard_1. \dots, \vard_{s}$ the \emph{subjects} of these atoms.
A non-constant atom is said to be \emph{nontrivial} if its coefficient is \emph{not} identical to zero.

In the rest of this subsection, we assume that the transition graph of $\Ss$ comprises a handle $H=q_0 \xrightarrow{(g_1,\eta_1)} q_1 \dots q_{m-1} \xrightarrow{(g_m,\eta_m)} q_{m}$, a collection of simple cycles $C_1,\dots,C_n$ such that $q_m$ is the unique state shared by each pair of distinct cycles from $\{C_1,\dots,C_n\}$, and a $\triangleright$-transition $q_m \xrightarrow{(\cur = \triangleright, \eta)} q_{m+1}$. Moreover, without loss of generality, we assume that $O(q_{m+1}) = a_0 + a_1 x_1 + \dots + a_k x_k + b_1 y_1 + \dots + b_l y_l$, and $O(q)$ is undefined for all the other states $q$. For convenience, we define $O(q_m)$ by replacing simultaneously $z$ with $\eta(z)$ in $O(q_{m+1})$,  for each $z \in \dom(\eta)$. Suppose $O(q_m) = a'_0 + a'_1 x_1 + \dots + a'_k x_k + b'_1 y_1 + \dots + b'_l y_l$. Then for the non-zero output problem, we can ignore the $\triangleright$-transition and use $O(q_m)$  directly.

A \emph{cycle scheme} $\schm$ is a path $C_{i_1}^{\ell_1} C_{i_2}^{\ell_2} \dots C_{i_t}^{\ell_t}$ such that $i_1,\dots,i_t \in [n]$, $\ell_1,\dots, \ell_t \ge 1$, and for each $j\in [t-1]$, $i_j \neq i_{j+1}$. Intuitively, $\schm$ is a path obtained by first iterating $C_{i_1}$ for $\ell_1$ times, then $C_{i_2}$ for $\ell_2$ times, and so on. From Proposition~\ref{prop-sum-cycle} and Corollary~\ref{cor-comp-two-paths}, a symbolic valuation $\sumf^{(\schm,\initval)}$ can be constructed 
to summarize the computation of $\Ss$ on $\schm$.

\begin{lemma}\label{prop-cycle-schm}
Suppose $\schm=C_{i_1}^{\ell_1} C_{i_2}^{\ell_2} \dots C_{i_t}^{\ell_t}$ is a cycle scheme, and $\initval$ is a symbolic valuation representing the initial values of the control and data variables such that for each $x_i, x_j \in X$, $\initval(x_i) = \initval(x_j)$ iff $i \sim_{q_m} j$. 
For all $j' \in  I^{\circled{C_{i_{1}}}}_{pe} \cap \rng(\pi^{\circled{q_m}})$, let $r_{j'}$ be the largest number $r \in [t]$ such that $j'\in\bigcap_{s\in[r]} I^{\circled{C_{i_{s}}}}_{pe}$, i.e., $x_{j'}$ remains persistent when traversing $C_{i_1}^{\ell_1} C_{i_2}^{\ell_2} \dots C_{i_{r_{j'}}}^{\ell_{r_{j'}}}$.
Then for each $j\in [l]$ and $j' \in I^{\circled{C_{i_1}}}_{pe}  \cap \rng(\pi^{\circled{q_m}})$, the coefficient of the $\initval(x_{j'})$-atom in $\sumf^{(\schm,\initval)}(y_j)$ is 
\begin{center}
\resizebox{0.8\hsize}{!}{
$e+\sum\limits_{s_1\in[r_{j'}]}  
\left(1+\lambda^{\circled{C_{i_{s_1}}}}_{j} + \dots + (\lambda^{\circled{C_{i_{s_1}}}}_{j})^{\ell_{s_1}-1} \right) \csta^{\circled{C_{i_{s_1}}}}_{j,(\pi^{\circled{q_m}})^{-1}(j')}\prod\limits_{{s_2}\in[{s_1}+1,t]}\left(\lambda^{\circled{C_{i_{s_2}}}}_{j}\right)^{\ell_{s_2}}$},
\end{center}
where (1) $e\!=\!0$ when $r_{j'}\!=\!t$ and (2) $e=(\lambda^{\circled{C_{i_s}}}_{j})^{\ell_s-1} \csta^{\circled{C_{i_{s}}}}_{j, (\pi^{\circled{q_m}})^{-1}(j')} \prod\limits_{{s'}\in[s+1,t]}\left(\lambda^{\circled{C_{i_{s'}}}}_{j}\right)^{\ell_{s'}}$ with $s=r_{j'}+1$ when $r_{j'}<t$.\\
The constant atom of $\sumf^{(\schm,\initval)}(y_j)$ is 
\begin{center}
\resizebox{0.7\hsize}{!}{$
\sum\limits_{{s_1}\in[t]}
\left(1+\lambda^{\circled{C_{i_{s_1}}}}_{j} + \dots + (\lambda^{\circled{C_{i_{s_1}}}}_{j})^{\ell_{s_1}-1} \right)
\cste^{\circled{C_{i_{s_1}}}}_{j} 
\prod\limits_{{s_2}\in[{s_1}+1,t]}\left(\lambda^{\circled{C_{i_{s_2}}}}_{j}\right)^{\ell_{s_2}}$}
\end{center}
Moreover, for all $j\!\in\! [l]$, in $\sumf^{(\schm,\initval)}(y_j)$, only the constant atom and the coefficients of the $\initval(x_{j'})$-atoms with $j' \!\in\! I^{\circled{C_{i_1}}}_{pe} \cap \rng(\pi^{\circled{q_m}})$ contain a subexpression of the form $ \mu_\schm \ell_1$ for some~$\mu_\schm\in \intnum$.
\end{lemma}
Notice that above, $\lambda^{\circled{C_{i_{s_1}}}}_j\in\{0,1\}$ for $j\in[l]$ and $s_1\in [t]$. Hence the value of $(1+\lambda^{\circled{C_{i_{s_1}}}}_{j} + \dots + (\lambda^{\circled{C_{i_{s_1}}}}_{j})^{\ell_{s_1}-1} )$ can only be $1$ or $\ell_{s_1}$ and $\left(\lambda^{\circled{C_{i_{s_2}}}}_{j}\right)^{\ell_{s_2}}\in\{0,1\}$.
Therefore, both the constant atom and the coefficients of the $\initval(x_{j'})$-atoms with $j' \!\in\! I^{\circled{C_{i_1}}}_{pe} \cap \rng(\pi^{\circled{q_m}})$ can be rewritten to the form of $c_0+c_1\ell_1+c_2\ell_2+\dots+c_t\ell_t$ for $c_0\ldots c_t\in \intnum$. Note that some of $c_0\ldots c_t$ might be zero.

\subsubsection{Step I:} 
We are ready to present the decision procedure. At first, we observe that  after traversing $H$ with the initial values of the variables given by some valuation $\sval_0$, for each $j' \in I^{\circled{H}}_{tr}$, the value of the control variable $x_{j'}$ becomes $\vard^{\circled{H}}_{\pi^{\circled{H}}_{tr}(j')}$,  more formally, $\sumf^{(H,\sval_0)}(x_{j'})=\vard^{\circled{H}}_{\pi^{\circled{H}}_{tr}(j')}$.

In Step I, we check if $\eval{O(q_m)}{\sumf^{(H,\sval_0)}}$ is not identical to zero.
This can be done by checking if the constant-atom or the coefficient of some non-constant atom of the output expression $\eval{O(q_m)}{\sumf^{(H,\sval_0)}}$ is not identical to zero.
\bigskip\\
\framebox[\textwidth]{
\begin{minipage}{0.95\textwidth}
\noindent {\bf Step I}. Decide whether $\eval{O(q_m)}{\sumf^{(H,\sval_0)}}$ is not identical to zero.
If the answer is yes, then the decision procedure terminates and returns the answer $\ltrue$. Otherwise, go to Step II.
\end{minipage}
}\medskip

\noindent{\it Complexity analysis of Step I}. Since $\sumf^{(H,\sval_0)}$ can be computed in polynomial time from $H$, it follows that Step I can be done in polynomial time.

\begin{example}\label{exmp-step-1}
Let $\Ss'_{\sf max}$ be the SNT in Example~\ref{exmp-sum}. Since $q_1 \xrightarrow{(\cur = \triangleright, \emptyset)} q_2$, we can define $O(q_1):= a'_0 +  a'_1 x_1 + b'_1 y_1 + b'_2 y_2 + b'_4y_3 = O(q_2) = y_1 - 2y_2 + y_3$.  The handle $H$ comprises exactly one transition. Thus $r^{\circled{H}}=1$ and 
\[\eval{O(q_1)}{\sumf^{(H, \initval_0)}} = \vard^{\circled{H}}_1 - 2 \vard^{\circled{H}}_1 + \vard^{\circled{H}}_1=0.\]
Therefore, the handle $H$ does not produce a non-zero output.
\end{example}

\subsubsection{Step II:}
The goal of Step II is to show that 
\begin{itemize}
\item either for each cycle scheme $\schm$, all subexpressions in $\eval{O(q_{m})}{\sumf^{(\schm,\sumf^{(H,\sval_0)})}}$ containing the cycle counter variables are identical to zero and hence can be ignored, 
\item or  there exists a cycle scheme $\schm$ such that $\eval{O(q_{m})}{\sumf^{(\schm,\sumf^{(H,\sval_0)})}}$ is not identical to zero. 
\end{itemize}
Let $\schm=C_{i_1}^{\ell_1} C_{i_2}^{\ell_2} \dots C_{i_t}^{\ell_t}$ be a cycle scheme. From Lemma~\ref{prop-cycle-schm}, for each $j'\in I^{\circled{C_{i_1}}}_{pe} \cap \rng(\pi^{\circled{q_m}})$ and symbolic valuation $\sval$, the only subexpression containing $\ell_1$ in the coefficient of $\initval(x_{j'})$-atom of $\eval{O(q_{m})}{\sumf^{(\schm,\initval)}}$ is
\begin{center}
	\resizebox{0.7\hsize}{!}{$
\sum \limits_{1 \le j \le l} 
b'_j \left((\cstl^{\circled{C_{i_2}}}_{j})^{\ell_2} \dots (\cstl^{\circled{C_{i_t}}}_{j})^{\ell_t}\right) 
\left(1+\cstl^{\circled{C_{i_1}}}_{j} + \dots + (\cstl^{\circled{C_{i_1}}}_{j})^{\ell_1-1} \right) \csta^{\circled{C_{i_1}}}_{j, (\pi^{\circled{q_m}})^{-1}(j')}.
\hspace{4mm} (\ast)
$}
\end{center}
Since $\cstl^{\circled{C_{i_1}}}_{j}, \cstl^{\circled{C_{i_2}}}_{j}, \dots, \cstl^{\circled{C_{i_t}}}_{j} \in \{0, 1\}$, the expression $(\ast)$  can be rewritten as  
 $\mu_{\schm, (i_1,j')} \ell_1 + \nu_{\schm, (i_1,j')}$ for some integer constants $\mu_{\schm, (i_1,j')}$ and $\nu_{\schm, (i_1,j')}$. 
 
The only subexpression containing $\ell_1$ in the constant atom of  $\eval{O(q_{m})}{\sumf^{(\schm,\initval)}}$ is
\begin{center}
	\resizebox{0.7\hsize}{!}{$
\sum \limits_{1 \le j \le l} b'_j
\begin{array}{l}
 \left((\lambda^{\circled{C_{i_2}}}_{j})^{\ell_2} \dots (\lambda^{\circled{C_{i_t}}}_{j})^{\ell_t}\right)
\left(1+\lambda^{\circled{C_{i_1}}}_{j} + \dots + (\lambda^{\circled{C_{i_1}}}_{j})^{\ell_1-1} \right) \cste^{\circled{C_{i_1}}}_{j}. \hspace{2mm} (\ast\ast)
\end{array}
$}
\end{center}
The expression $(\ast\ast)$ can be rewritten as $\mu_{\schm,(i_1,0)} \ell_1 + \nu_{\schm,(i_1,0)}$ for some integer constants $\mu_{\schm, (i_1,0)}$ and $\nu_{\schm, (i_1,0)}$.

\smallskip
We are ready to present Step II.
\bigskip\\
\framebox[\textwidth]{
	\begin{minipage}{0.95\textwidth}
		\noindent {\bf Step II}. For each $i_1 \in [n]$, check all cycle scheme $\schm=C_{i_1}^{\ell_1} C_{i_2} \dots C_{i_t}$ such that $i_2,\dots,i_t$ are mutually distinct. There are only finitely many this kind of cycle schemes. If 
		one of the following constraints is satisfied, then return $\ltrue$. 
		\begin{enumerate}
		\item There is  $ j' \in I^{\circled{C_{i_1}}}_{pe} \cap \rng(\pi^{\circled{q_m}})$ such that $ \mu_{\schm,(i_1,j')} \neq 0$. 
		\item $\mu_{\schm,(i_1,0)} \neq 0$.
		\end{enumerate}
		If the decision procedure has not returned yet, then go to Step III.
	\end{minipage}
}\bigskip

\noindent{\it Complexity analysis of Step II}. Since $i_1,\dots, i_t$ are mutually distinct, the number of cycle schemes $\schm = C_{i_1}^{\ell_1} C_{i_2} \dots C_{i_t}$ in Step II is exponential in the number of cycles in the generalized lasso. Once the cycle scheme $\schm$ is fixed, the two constraints in Step II can be decided in polynomial time. Therefore, the complexity of Step II is exponential in the number of simple cycles in the generalized lasso.

\begin{example}\label{exmp-step-2}
Let $\Ss'_{\sf max}$ be the SNT in Example~\ref{exmp-step-1}. We need consider the following cycle schemes, $C^{\ell_1}_1$, $C^{\ell_1}_2$, $C^{\ell_1}_1 C_2$, $C^{\ell_1}_2 C_1$, $C^{\ell_1}_1 C_2 C_1$, $C^{\ell_1}_2 C_1 C_2$. We use $\schm_1 = C^{\ell_1}_1$, $\schm_2 = C^{\ell_1}_1 C_2$, and $\schm_3 = C^{\ell_1}_1 C_2 C_1$ to illustrate Step II.
\begin{itemize}
\item In $\eval{O(q_1)}{\sumf^{(\schm_1, \initval)}}$,  the subexpression containing $\ell_1$ in the coefficient of the $\initval(x_1)$-atom is
\[
\begin{array} {l}
\sum \limits_{1 \le j \le 3} b'_j
\begin{array}{l}
\left(1+\lambda^{\circled{C_{1}}}_{j} + \dots + (\lambda^{\circled{C_{1}}}_{j})^{\ell_1-1} \right) \csta^{\circled{C_{1}}}_{j, 1}
\end{array}
\\
= 1 \times \ell_1 \times 4 +  (-2) \times \ell_1 \times 2 + 1 \times 1 \times 0 = 0.
\end{array}
\]
\item In $\eval{O(q_1)}{\sumf^{(\schm_2, \initval)}}$, the subexpression containing $\ell_1$ in the coefficient of the $\initval(x_1)$-atom is 
\[
\begin{array}{l}
\sum \limits_{1 \le j \le 3} b'_j
\begin{array}{l}
 \lambda^{\circled{C_{2}}}_{j}
\left(1+\lambda^{\circled{C_{1}}}_{j} + \dots + (\lambda^{\circled{C_{1}}}_{j})^{\ell_1-1} \right) \csta^{\circled{C_{1}}}_{j, 1}
\end{array}
\\
= 1 \times 1 \times \ell_1 \times 4 + (-2) \times 1 \times \ell_1 \times 2 + 1 \times 1 \times 1 \times 0 =0.
\end{array}
\]
\item In $\eval{O(q_1)}{\sumf^{(\schm_3, \initval)}}$, the subexpression containing $\ell_1$ in the coefficient of the $\initval(x_1)$-atom is 
\[
\begin{array}{l}
\sum \limits_{1 \le j \le 3} b_j
\begin{array}{l}
 (\lambda^{\circled{C_{2}}}_{j}\lambda^{\circled{C_{1}}}_{j})
\left(1+\lambda^{\circled{C_{1}}}_{j} + \dots + (\lambda^{\circled{C_{1}}}_{j})^{\ell_1-1} \right) \csta^{\circled{C_{1}}}_{j, 1}
\end{array}
\\
= 1 \times 1 \times \ell_1 \times 4 + (-2) \times 1 \times \ell_1 \times 2 + 1 \times 0 \times 1 \times 0 =0.
\end{array}
\]
\end{itemize}
From Example~\ref{exmp-sum}, we know that if $\schm=C^{\ell_1}_2, C^{\ell_1}_2 C_1$, or $C^{\ell_1}_2 C_1 C_2$, then for each $j =1, 2,3$, the coefficient of the $\initval(x_1)$-atom in $\sumf^{(\schm, \initval)}(y_j)$ does not contain the cycle counter variable $\ell_1$. Therefore, the coefficient of the $\initval(x_1)$-atom in $\eval{O(q_1)}{\sumf^{(\schm, \initval)}}$ does not contain $\ell_1$ as well. From this, we conclude that with $\Ss'_{\sf max}$ as the input, the decision procedure does not return in Step II and it will go to Step III.
\end{example}

\medskip
In the following, we present the arguments for the correctness of Step II. Suppose $\schm =C_{i_1}^{\ell_1} C_{i_2} \dots C_{i_t}$.
\begin{itemize}
\item If there exists $j' \in I^{\circled{C_{i_1}}}_{pe} \cap \rng(\pi^{\circled{q_m}})$ and 
$\mu_{\schm,(i_1,j')} \neq 0$. 
Then from the cycle scheme $\schm=C_{i_1}^{\ell_1} C_{i_2} \dots C_{i_t}$, we can assign a sufficiently large value $s$ to $\ell_1$ so that the coefficient of the $\initval(x_{j'})$-atom in $\eval{O(q_{m})}{\sumf^{(H\schm ,\sval_0)}}$, which is equal to the sum of  $\ell_1 \mu_{\schm,(i_1,j')}$ and some constant,  becomes non-zero.  The guards and assignments in the path $H C_{i_1}^{s} C_{i_2} \dots C_{i_t}$ enforce a preorder over the subjects of those nontrivial non-constant atoms.
Pick one of the nontrivial non-constant atoms with a maximal subject w.r.t. the preorder. Since the subject is maximal, it can be assigned an arbitrarily large number so that the corresponding atom dominates $\eval{O(q_{m})}{\sumf^{(P\schm,\sval_0)}}$.  This is sufficient to make $\eval{O(q_{m})}{\sumf^{(H\schm,\sval_0)}}$ non-zero.
\item Otherwise, if $\eval{O(q_{m})}{\sumf^{(H\schm,\sval_0)}}$ contains some other nontrivial non-constant atoms, then we can apply a similar argument as above and conclude that $\eval{O(q_{m})}{\sumf^{(H\schm,\sval_0)}}$ can be made non-zero. 

\item On the other hand, if $\eval{O(q_{m})}{\sumf^{(H\schm,\sval_0)}}$ contains no nontrivial non-constant atoms, but $\mu_{\schm,(i_1,0)} \neq 0$, then we can let $\ell_1$ sufficiently large to make the expression $\eval{O(q_{m})}{\sumf^{(H\schm,\sval_0)}}$ non-zero. 
\end{itemize} 
Therefore, when Step II returns $\ltrue$, that is, at least one of the two conditions in Step II holds, then we are able to conclude that there \emph{must be}  an input data word $w$ and an initial valuation $\rho_0$ such that $ \Ss_{\rho_0}(w) \not \in \{\bot, 0\}$. 

\smallskip

If Step II does not return $\ltrue$, we show below that for all cycle schemes $\schm_1=C_{i_1}^{\ell_1} C_{i_2}^{\ell_2} \dots C_{i_{t_1}}^{\ell_{t_1}}$ with $i_1,i_2,\dots,i_{t_1} \in [n]$, all the subexpressions containing cycle counter variables $\ell_1,\dots, \ell_{t_1}$ in $\eval{O(q_{m})}{\sumf^{(\schm_1, \sumf^{(H,\sval_0)})}}$ are identical to zero and hence can be removed. Let ${i'_2} \dots {i'_{t_2}}$ be the sequence obtained from $i_2 \dots i_{t_1}$ by keeping just one copy for each duplicated index therein.  
In Step II we already checked the cycle scheme $\schm_2=C_{i_1}^{\ell_1} C_{i'_2} \dots C_{i'_{t_2}}$. Step II guarantees that all the subexpressions containing $\ell_1$ in 
$\eval{O(q_{m})}{\sumf^{(\schm_2, \sumf^{(H,\sval_0)})}}$ are identical to zero and hence can be removed.
Because for all $j\in[l]$, $\cstl^{^{\circled{C_1}}}_j, \dots, \cstl^{^{\circled{C_n}}}_j \in \{0,1\}$,   $(\lambda^{\circled{C_{i_2}}}_{j})^{\ell_2} \dots (\lambda^{\circled{C_{i_{t_1}}}}_{j})^{\ell_{t_1}} = \lambda^{\circled{C_{i'_2}}}_{j} \dots \lambda^{\circled{C_{i'_{t_2}}}}_{j}$. We proved that the $(\ast)$ and $(\ast\ast)$ style expressions are equivalent in both $\eval{O(q_{m})}{\sumf^{(\schm_1, \sumf^{(H,\sval_0)}))}}$ and $\eval{O(q_{m})}{\sumf^{(\schm_2, \sumf^{(H,\sval_0)}))}}$.
Hence we can also remove all subexpressions containing $\ell_1$ from  $\eval{O(q_{m})}{\sumf^{(\schm_1, \sumf^{(H,\sval_0)}))}}$, without affecting its value.
Those subexpressions containing $\ell_2$ can also be removed by considering the cycle scheme $\schm_3=C_{i_2}^{\ell_2} C_{i''_3} \dots C_{i''_{t_3}}$ and applying a similar reasoning, where the sequence ${i''_3} \dots {i''_{t_3}}$ is obtained from ${i_3} \dots  i_{t_1}$, similarly to the construction of ${i'_2} \dots {i'_{t_2}}$ from $i_2 \dots i_{t_1}$. The same applies to all other cycle counter variables $\ell_3,\dots,\ell_{t_1}$.
For each $y_j \in Y$, we use the notation ${\sumf^{(\schm_1,\initval)}}^-(y_j)$ to denote the expression obtained by removing from the constant atom and coefficients of the non-constant atoms of $\sumf^{(\schm_1,\initval)}(y_j)$ all subexpressions containing the cycle counter variables. 

\begin{lemma}\label{prop-bnd-domain-1}
	Suppose that the decision procedure has not returned $\ltrue$ after Step~II. For each cycle scheme $\schm$, let $f=\eval{O(q_{m})}{\sumf^{(\schm, \sumf^{(H,\sval_0)})}}$ and $f'=\eval{O(q_{m})}{{\sumf^{(\schm, \sumf^{(H,\sval_0)})}}^-}$. For all valuations $\rho$, $\eval{f}{\rho}\neq 0$ iff $\eval{f'}{\rho} \neq 0$.
\end{lemma}

\subsubsection{Step III:} 

From Proposition~\ref{prop-sum-cycle},  we can observe that for each cycle $C_i$ (where $1 \le i \le n$),  the expression $\sumf^{(C^\ell_i, \sumf^{(H,\sval_0)})^-}(y_j)$ for $y_j \in Y$ is of one of the the following forms.
\begin{itemize}
\item If $\cstl^{\circled{C_i}}_{j}=0$, then $\sumf^{(C^\ell_i, \sumf^{(H,\sval_0)})^{-}}(y_j) = \sumf^{(C^\ell_i, \sumf^{(H,\sval_0)})}(y_j)$.

\item If $\cstl^{\circled{C_i}}_{j}=1$, then
\medskip\\
\resizebox{0.95\hsize}{!}{$
\begin{array}{l c l}
\sumf^{(C^\ell_i, \sumf^{(H,\sval_0)})^-}(y_j)  &=&    \sumf^{(H,\sval_0)}(y_j) +   
\sum \limits_{j' \in [s^{\circled{q_m}}] \setminus \rng(\pi^{\circled{C_i}}_{pe})} \csta^{\circled{C_i}}_{j,j'} \initval(x_{\pi^{\circled{q_m}}(j')})\ +   \\
\medskip
& & \sum \limits_{j' \in \rng(\pi^{\circled{C_i}}_{tr})} \sum \limits_{s\in[\ell -1]}
\left(\cstb^{\circled{C_i}}_{j,j'} + \sum \limits_{j'' \in (\pi^{\circled{C_i}}_{tr})^{-1}(j') \cap \rng(\pi^{\circled{q_m}})} \csta^{\circled{C_i}}_{j, (\pi^{\circled{q_m}})^{-1}(j'')}  \right) \vard^{\circled{C_i , s}}_{j'}\ + \\
\medskip
& & \sum \limits_{j' \in [r^{\circled{C_i}}]  \setminus \rng(\pi^{\circled{C_i}}_{tr}) }\sum \limits_{s\in[\ell -1]} 
\cstb^{\circled{C_i}}_{j,j'} \vard^{\circled{C_i , s}}_{j'} + \sum \limits_{j' \in [r^{\circled{C_i}}] }  
\cstb^{\circled{C_i}}_{j, j'} \vard^{\circled{C_i , \ell}}_{j'}.
\end{array}
$}
\end{itemize}
Note that if $\cstl^{\circled{C_i}}_{j}=1$, then the expressions containing the cycle counter variable $\ell$, e.g. $\ell \cste^{\circled{C_i}}_j$, are removed from $\sumf^{(C^\ell_i, \sumf^{(H,\sval_0)})^-}(y_j)$.

We define the abstraction of $\sumf^{(\schm,\sumf^{(H,\initval_0)})^-}$, denoted by $\abs(H\schm)$, as the union of the following three sets of tuples:
\begin{itemize}
\item the tuple for the constant atom: $\left\{\left(0, \left( \cste^{(\circled{\schm})^-}_{1}+ \cstl^{\circled{\schm}}_{1} \cste^{\circled{H}}_{1},\dots, \cste^{(\circled{\schm})^-}_{l}+ \cstl^{\circled{\schm}}_{l} \cste^{\circled{H}}_{l} \right) \right)\right\}$,
\item tuples for the control variable atoms: $\{(j', (c_{j',1},\dots, c_{j', l})) \mid j' \in [k]\}$, where $c_{j', j}$ is the coefficient of the ${\sumf^{(\schm,\sumf^{(H,\sval_0)})}}(x_{j'})$-atom in ${\sumf^{(\schm,\sumf^{(H,\sval_0)})}}^-(y_{j})$ for $j\in[l]$,
\item tuples for the other atoms: $\{(k+1, (c_1,\dots,c_l))\}$, where $(c_1,\dots,c_l)$ is the vector of coefficients of the $\vard'$-atom in ${(\sumf^{(\schm,\sumf^{(H,\sval_0)})}}^-(y_{j})$ for all $j \in [l]$ and $\vard'\not\in \{{\sumf^{(\schm,\sumf^{(H,\sval_0)})}}(x_{j'})\mid x_{j'}\in X\}$.
\end{itemize}
Note that in $\abs(H\schm)$, if $j_1, j_2 \in [k]$ and $j_1 \sim_{q_m} j_2$, then we have $\sumf^{(\schm,\sumf^{(H,\initval_0)})}(x_{j_1})=\sumf^{(\schm,\sumf^{(H,\initval_0)})}(x_{j_2})$. Therefore, $(c_{j_1, 1}, \dots, c_{j_1, l}) = (c_{j_2,1}, \dots, c_{j_2, l})$.

Let $\absset=\bigcup \{\abs(H\schm) \mid \schm \mbox{ a cycle scheme}\}$. Then $\absset$ can be constructed inductively as follows, until $\absset_{i+1}= \absset_i$. 
\begin{enumerate}
\item $\absset_0 = \{\abs(HC_1), \ldots \abs(HC_n)\}$,

\item For $i \ge 0$, $\absset_{i+1}$ is the union of $\absset_i$ and the set of $\Lambda'$ such that $\Lambda'$ is constructed from some $\Lambda \in \absset_i$ and some cycle $C_{i'}$ (where $ i' \in [n]$) as follows. At first we observe that $\sim_{q_m} \subseteq (I^{\circled{C_{i'}}}_{pe} \times I^{\circled{C_{i'}}}_{pe}) \cup (I^{\circled{C_{i'}}}_{tr} \times I^{\circled{C_{i'}}}_{tr})$. The argument is as follows: If $j' \in I^{\circled{C_{i'}}}_{tr}$ and $j'' \in I^{\circled{C_{i'}}}_{pe}$, then $x_{j'}$ is assigned a fresh value and $x_{j''}$ is assigned the initial value of some control variable, thus $x_{j'}$ and $x_{j''}$ are not equivalent w.r.t. $\sim_{q_m}$.
%
\begin{itemize}
\item Suppose $(0, (c_1, \dots, c_l)) \in \Lambda$. Then $(0, (c'_1, \dots, c'_l)) \in \Lambda'$, where for each $j \in [l]$, if $\cstl^{\circled{C_{i'}}}_j = 0$, then $c'_j = \cste^{\circled{C_{i'}}}_{j}$, otherwise, $c'_j = c_j$ (in this case, the expression $\cste^{\circled{C_{i'}}}_{j}$ is removed).  
\item For each $j' \in [k]$, suppose $(j', (c_{j',1}, \dots, c_{j',l})) \in \Lambda$, do the following. 
\begin{itemize}
\item If $j' \in I^{\circled{C_{i'}}}_{pe}$, then $(j', (c'_{j', 1}, \dots, c'_{j', l})) \in \Lambda'$,  where for each $j \in [l]$, 
\begin{itemize}
\item if $\cstl^{\circled{C_{i'}}}_j = 0$, then $c'_{j', j}=  \csta^{\circled{C_{i'}}}_{j, \pi^{\circled{C_{i'}}}_{pe}(j')}$, 
\item otherwise, $c'_{j', j} = c_{j',j} $ (in this case, the expressions $\csta^{\circled{C_{i'}}}_{j, \pi^{\circled{C_{i'}}}_{pe}(j')}$ is removed).
\end{itemize}
\item If $j' \in I^{\circled{C_{i'}}}_{tr}$, then 
$$(k+1, (c'_{j', 1}, \dots, c'_{j', l})), \left(j', \left(\cstb^{\circled{C_{i'}}}_{1, \pi^{\circled{C_{i'}}}_{tr}(j')}, \dots, \cstb^{\circled{C_{i'}}}_{l, \pi^{\circled{C_{i'}}}_{tr}(j')} \right) \right) \in \Lambda',$$ 
where for each $j \in [l]$, if $\cstl^{\circled{C_{i'}}}_j = 0$, then $c'_{j', j}= \csta^{\circled{C_{i'}}}_{j, (\pi^{\circled{q_m}})^{-1}(j''_0)}$,  otherwise, $c'_{j', j} = c_{j',j} +   \csta^{\circled{C_{i'}}}_{j, (\pi^{\circled{q_m}})^{-1}(j''_0)}$, where $j''_0 = \min(\{j'' \in [k] \mid j'' \sim_{q_m} j' \})$. In this case, after going through $C_{i'}$, the control variable $x_{j'}$ stores a fresh value and the initial value of $x_{j'}$ is not stored in any control variable, thus the $(j',\dots)$-tuple is updated and a $(k+1,\dots)$-tuple is added for the initial value of $x_{j'}$.
\end{itemize}

\item For each tuple $(k+1, (c_1, \dots, c_l)) \in \Lambda$, we have $(k+1, (c'_1, \dots, c'_l)) \in \Lambda'$, where for each $j \in [l]$, if $\cstl^{\circled{C_{i'}}}_j = 0$, then $c'_{j}= 0$, otherwise, $c'_{j} = c_j$. In addition, for each $j' \in [r^{\circled{C_{i'}}}] \setminus \rng(\pi^{\circled{C_{i'}}}_{tr})$, $(k+1, (\cstb^{\circled{C_{i'}}}_{1, j'}, \dots, \cstb^{\circled{C_{i'}}}_{l, j'})) \in \Lambda'$. 
\end{itemize}
\end{enumerate}

By a simple analysis on the inductive computation of $\absset$, we can show that all the tuples $(j', (c_1, \dots, c_l))$ occurring in $\absset$ satisfy that $c_1,\dots, c_l$ are from a bounded domain $U$, which is stated in the following lemma.

\begin{lemma}\label{prop-bnd-domain-2}
	Suppose that the decision procedure has not returned yet after Step II. 
	For all cycle scheme $\schm$ and $y_j \in Y$, the constant atom and the coefficients of all non-constant atoms in ${\sumf^{(\schm, \sumf^{(H,\initval_0)})}}^-(y_j)$ are from a finite set $U \subset \intnum$ comprising
the constant atom and the coefficients of the non-constant atoms in the expression $\sumf^{(C^{\ell_{i_1}}_{i_1}, \sumf^{(H, \initval_0)})^-}(y_j)$ or $\sumf^{(C_{i_1}C_{i_2}, \sumf^{(H, \initval_0)})^-}(y_j)$ for $i_1, i_2\in [n]$ such that $i_1 \neq i_2$ and $\ell_{i_1} \in \{1,2\}$.
\hide{
	\item the numbers  $2\sum \limits_{j'' \sim_{q_m} j'} \csta^{\circled{C_{i'}}}_{j,j''}$ for $i' \in [n]$ such that $\cstl^{\circled{C_{i'}}}_j = 0$, $i'' \in [n]$ such that $\cstl^{\circled{C_{i''}}}_j = 1$, and $j' \in I^{\circled{C_{i'}}}_{pe} \cap I^{\circled{C_{i''}}}_{tr}$, 
	\item the numbers $\cstb^{\circled{H}}_{j,\pi^{\circled{H}}_{tr}(j')} + \sum \limits_{j'' \sim_{q_m} j'} \csta^{\circled{C_{i'}}}_{j,j''}$ for $i' \in [n]$ and $j' \in I^{\circled{H}}_{tr} \cap I^{\circled{C_{i'}}}_{tr}$, 

	\item the numbers $\sum \limits_{j'' \sim_{q_0} \pi^{\circled{H}}_{pe}(j')} \csta^{\circled{H}}_{j, j''} + \sum \limits_{ j''' \sim_{q_m} j' } \csta^{\circled{C_{i'}}}_{j,j'''}$ for $i' \in [n]$ and $j' \in I^{\circled{H}}_{pe} \cap I^{\circled{C_{i'}}}_{tr}$, 
	\item the numbers $\cstb^{\circled{C_i}}_{j,\pi^{\circled{C_i}}(j')} + \sum \limits_{j'' \in J} \csta^{\circled{C_{i'_{j''}}}}_{j,j''}$ for $j' \in I^{\circled{C_i}}_{tr}$, $J \subseteq \{j'' \in I^{\circled{C_i}}_{tr} \mid  \pi^{\circled{C_i}}_{tr}(j'') = \pi^{\circled{C_i}}_{tr}(j')\}$, and $i'_{j''} \in [n]$  for each $j'' \in J$. 
}
\end{lemma}

Note that although the set $U$ can be exhausted by the cycle schemes stated in Lemma~\ref{prop-bnd-domain-2}, the inductive computation of $\absset$ may not be.
\bigskip\\
\framebox[\textwidth]{
	\begin{minipage}{0.95\textwidth}
		\noindent {\bf Step III.} We first construct the set $\absset$. Then for each $\Lambda \in \absset$, do the following.
		\begin{enumerate}
			\item If $(0,(c_{0,1},\dots,c_{0,l})) \in \Lambda$ such that $a'_0+b'_1 c_{0,1}+\dots + b'_l c_{0,l} \neq 0$, then return $\ltrue$.
			\item If there is $j' \in [k]$ such that $(\sum \limits_{j'' \sim_{q_m} j'} a'_{j''}) + b'_1 c_{j',1} + \dots + b'_l c_{j',l} \neq 0$, then  return $\ltrue$, where $(j', (c_{j',1},\dots,c_{j',l})) \in \Lambda$.
			\item If there is $(k+1,(c_1,\dots,c_l)) \in \Lambda$ such that $b'_1 c_1 + \dots + b'_l c_l \neq 0$, then return $\ltrue$. 
		\end{enumerate}
		If the decision procedure has not returned yet, return $\lfalse$.
	\end{minipage}
}\medskip\\

In order to reduce the size of $\absset$, we can restructure $\absset$ into a pair $\absset'=(\Xi, \Delta)$ as follows, without affecting the computation in Step III.
\begin{enumerate}
\item Initially, let $\Xi = \Delta := \emptyset$.
\item For each $\Lambda \in \absset$, do the following: Let $\Lambda' := \Lambda \setminus \{(k+1, (c_1, \dots, c_l)) \in \Lambda\}$. In addition, let $\Xi := \Xi \cup \{\Lambda'\}$ and $\Delta:= \Delta \cup \{(k+1, (c_1,\dots, c_l)) \in \Lambda\}$. 
\end{enumerate}

\noindent {\it Complexity analysis of Step III}. The size of the set $U$ is polynomial in the size of generalized lasso. Then size of $\Xi$ is at most exponential in $kl$ and the size of $\Delta$ is at most exponential in $l$. Therefore, the size of $\absset'$ is at most exponential in $kl$ and the computation of $\absset'$ takes exponential time in the worst case. The three conditions in Step III can be checked in time polynomial over the size of $\absset'$. In summary, the complexity of Step III is exponential in $kl$, the product of the number of control and data variables.

\begin{example}\label{exmp-step-3}
Let $\Ss'_{\sf max}$ be the SNT in Example~\ref{exmp-step-2}. Then
\[
\begin{array}{l c l}
\sumf^{(C^\ell_1, \initval)^{-}}(y_1) &= & \initval(y_1) + 0 \initval(x_1) +\ \ \vard^{\circled{C_1, 1}}_1 + \dots +\ \ \vard^{\circled{C_1, \ell}}_1,\\
\sumf^{(C^\ell_1, \initval)^-}(y_2) &=& \initval(y_2) + 0 \initval(x_1) + 2 \vard^{\circled{C_1, 1}}_1 + \dots + 2 \vard^{\circled{C_1, \ell}}_1,\\
\sumf^{(C^\ell_1, \initval)^-}(y_3) &=& \hspace{5.1cm} 3 \vard^{\circled{C_1, \ell}}_1,
\end{array}
\] 
and $\sumf^{(C^\ell_2, \initval)^-}(y_j) = \sumf^{(C^\ell_2, \initval)}(y_j)$ for each $j=1,2,3$.
The computation of $\absset$ starts with the set $\absset_0 = \{\abs(HC_1), \abs(HC_2)\}$. We illustrate how to compute $\abs(HC_1)$ and $\abs(HC_2)$. 
\begin{itemize}
\item Since none of $\sumf^{(H, \initval_0)}(y_j)$'s , $\sumf^{(C^\ell_1, \initval)^{-}}(y_j)$'s and $\sumf^{(C^\ell_2, \initval)^{-}}(y_j)$'s contains constant atoms, we know that $(0, (0,0,0))$ occurs in $\abs(HC_1)$ and  $\abs(HC_2)$. 
\item After going through $H$, each of $y_1,y_2,y_3$ is assigned the first data value, and after going through $HC_1$, $x_1$ holds the first data value. From $\sumf^{(C^\ell_1, \initval)^{-}}(y_j)$'s mentioned above, we know that each of $\sumf^{(C_1, \sumf^{(H, \initval_0)^{-}})}(y_1)$ and $\sumf^{(C_1, \sumf^{(H, \initval_0)^{-}})}(y_2)$ holds one copy of the first data value,  and $\sumf^{(C_1, \sumf^{(H, \initval_0)^{-}})}(y_3)$ contains no copies of the first data value. Therefore, $(1, (1,1,0))$ occurs in $\abs(HC_1)$. Similarly, since $x_1$ holds the second data value after going through $HC_2$, we have $(1, (3, 2, 1))$ occurs in $\abs(HC_2)$. 
\item In addition, by a simple calculation, we know that $\abs(HC_1)$ contains another tuple $(2,(1, 2, 3))$ and $\abs(HC_2)$ contains another tuple $(2, (2, 4, 6))$.
\end{itemize}
To summarize, we have $\abs(HC_1)=\{(0, (0,0,0)), (1, (1, 1, 0)), (2, (1, 2, 3))\}$ and $\abs(HC_2) = \{(0, (0, 0, 0)), (1, (3, 2,1)), (2, (2, 4, 6))\}$. Then starting from $\absset_0$, we compute $\absset_1 = \absset_0\ \cup\ \{\absset(HC_{i_1}C_{i_2}) \mid i_1, i_2 = 1, 2\}$, and so on. We illustrate how to compute $\abs(HC_1 C_2)$ from $\abs(HC_1)$. 
\begin{itemize}
\item Since $(0, (0,0,0))$ occurs in $\abs(HC_1)$ and $\cstl^{\circled{C_2}}_1 = \cstl^{\circled{C_2}}_2 = \cstl^{\circled{C_2}}_3 = 1$, we know that $(0, (0,0,0)) \in \abs(HC_1C_2)$.

\item From $1 \in I^{\circled{C_2}}_{tr}$ and $\cstl^{\circled{C_2}}_1 = \cstl^{\circled{C_2}}_2 = \cstl^{\circled{C_2}}_3 = 1$, we have $(2, (1 + \csta^{\circled{C_2}}_{1, 1}, 1+\csta^{\circled{C_2}}_{2, 1}, 0 + \csta^{\circled{C_2}}_{3, 1})) = (2, (1+1, 1+3, 0 +5))= (2, (2, 4, 5)) \in \abs(HC_1C_2)$. In addition, we have $(1, (\cstb^{\circled{C_2}}_{1, 1}, \dots, \cstb^{\circled{C_2}}_{3, 1}))=(1, (3, 2, 1)) \in \abs(HC_1C_2)$.

\item Since $(2, (2, 4, 6))$ occurs in $\abs(HC_1)$ and $\cstl^{\circled{C_2}}_1 = \cstl^{\circled{C_2}}_2 = \cstl^{\circled{C_2}}_3 = 1$, we have $(2, (2, 4, 6)) \in \abs(HC_1C_2)$. Moreover, because $[r^{\circled{C_2}}] \setminus \rng(\pi^{\circled{C_2}}_{tr})=\emptyset$, no other tuples are added into $\abs(HC_1C_2)$.
\end{itemize}
In summary, 
$$\abs(HC_1C_2)= \{(0,(0,0,0)), (1, (3,2,1)), (2, (2,4,5)), (2, (2, 4,6))\}.$$
From the fact that $(1, (1, 1, 0))$ occurs in $\abs(HC_1)$, we know that $a'_1 + b'_1 \times 1 + b'_2 \times 1 + b'_3 \times 0 = 0 + 1 \times 1 + (-2) \times 1 + 1 \times 0 = -1 \neq 0$. Therefore, Step III returns $\ltrue$ and a non-zero output can be produced by following the path $HC_1$.
\end{example}

\vspace{-4mm}

\subsection{Decision procedure for SNTs}\label{sec-gflat}

We generalize the decision procedure for the special case that the transition graphs of SNTs are generalized lassos to the full class of SNTs.
We  define a \emph{generalized multi-lasso} as $\gmlasso= H_1 (C_{1,1},\dots,C_{1,n_1}) H_2 (C_{2,1},\dots,C_{2,n_2}) \dots H_r (C_{r,1},\dots, C_{r, n_r}) H_{r+1}$ s.t. (1) for each $s\in [r]$, $H_s=q_{s,1} \xrightarrow{(g_2,\eta_2)} q_{s,2} \dots q_{s,m_s-1} \xrightarrow{(g_{m_s},\eta_{m_s})} q_{s,m_s}$ is a generalized lasso, $H_{r+1} = q_{r, m_r} \xrightarrow{(\cur = \triangleright, \eta')} q'$, (2) for $1 \leq s< s' \leq r$, $H_s (C_{s,1},\dots,C_{s, n_s})$ and $H_{s'} (C_{s', 1},\dots,C_{s', n_{s'}})$ are state-disjoint, except the case that when $s'=s+1$, $q_{s, m_s}=q_{s',1}$, and (3) $q_{1,1}=q_0$.

Since the transition graph of $\Ss$ can be seen as a finite collection of generalized multi-lassos, in the following, we shall present the decision procedure by showing how to decide the non-zero output problem for generalized multi-lassos. 

In the following, we fix a generalized multi-lasso 

\smallskip
\hspace{2mm} $\gmlasso= H_1 (C_{1,1},\dots,C_{1,n_1}) H_2 (C_{2,1},\dots,C_{2,n_2}) \dots H_r (C_{r,1},\dots, C_{r, n_r}) H_{r+1}$,\\
\smallskip
 and assume without loss of generality that $O(q')=a_0+a_1 x_1 + \dots + a_k x_k + b_1 y_1  + \dots + b_l y_l$ and $O(q'')$ is undefined for every other state $q''$ in $\gmlasso$. For convenience, we adapt the output function $O$ a bit to define $O(q_{r,m_r}): = a'_0+a'_1 x_1 + \dots + a'_k x_k + b'_1 y_1  + \dots + b'_l y_l$, where $a'_0+a'_1 x_1 + \dots + a'_k x_k + b'_1 y_1  + \dots + b'_l y_l$ is obtained from $O(q')$ by replacing simultaneously each $z \in \dom(\eta')$ with $\eta'(z)$.

\vspace{-2mm}

\subsubsection{Step I$'$:} We do the same analysis as in Step I for the path $H_1\dots H_r$.

\hide{\smallskip\\
\framebox[\textwidth]{
	\begin{minipage}{0.95\textwidth}
		\noindent {\bf Step I$'$}: We do the same analysis as in Step I for the path $H_1\dots H_r$.
	\end{minipage}
}\smallskip}

\vspace{-3mm}
\subsubsection{Step II$'$:}
Let $s\in [1,r-1]$. In order to analyze the set of cycles $\Cc=\{C_{s,1},\dots,C_{s,n_{s}}\}$, next we show how to summarize the effect of the path $H_{s+1}\dots H_r$ on the values of the variables in the state $q_{s,_{m_s}}$ by extending the output function and defining $O(q_{s,_{m_s}})$ (note that $q_{s,_{m_s}}$ is the unique state shared by all those cycles in $\Cc$).
Suppose that $\eval{O(q_{r,m_r})}{\sumf^{(H_{s+1}\dots H_{r}, \initval)}}=a''_0+a''_1 \initval(x_1)+ \dots + a''_k \initval(x_k) + b''_1 \initval(y_1) + \dots + b''_l \initval(y_l)+e$, where $\initval(x_1)\dots\initval(x_k)$ and $\initval(y_1)\ldots \initval(y_l)$ represent the values of $x_1\dots x_k$ and $y_1 \dots y_l$ in the state $q_{s, m_{s}}$, and $e$ is a linear combination of the variables that represent the data values introduced when traversing $H_{s+1}\dots H_r$. 
Then we let
$O(q_{s, m_{s}}):=a''_0+a''_1 x_1 + \dots + a''_k x_k + b''_1 y_1 + \dots + b''_l y_l$.
%
\bigskip\\
\framebox[\textwidth]{
	\begin{minipage}{0.95\textwidth}
\noindent {\bf Step II$'$}.  For each $s\in [r]$, $s'\in [n_s]$, and each cycle scheme $\schm = C^{\ell_1}_{s,s'} C_{i_{2}} \dots C_{i_{t}}$ such that $C_{i_{2}} \dots C_{i_{t}}\in \{C_{s, 1}, \dots, C_{s,n_s},\dots, C_{r,1}, \dots, C_{r,n_r}\}$ and $C_{i_{2}} \dots C_{i_{t}}$ are mutually distinct, we perform an analysis of the expression $\eval{ O(q_{s, m_{s}})} {\sumf^{(\schm,\sumf^{(H_1 \dots H_{s}, \initval_0)} ) } }$, in a way similar to Step II. If the decision procedure does not return during the analysis, then go to Step III$'$.
	\end{minipage}
}\medskip\\
Intuitively, in Step II$'$, during the analysis of the cycle scheme $\schm = C^{\ell_1}_{s,s'} C_{i_{2}} \dots C_{i_{t}}$, the effect of the paths $H_{s+1},  \dots,  H_r$ and the cycles $C_{i_{2}}, \dots, C_{i_{t}}$ on the coefficients of atoms which contain the cycle counter variable $\ell_1$, is described by the expressions  $\cstl^{\circled{H_{s+1}}}_j \dots \cstl^{\circled{H_{r}}}_j  \cstl^{\circled{C_{i_{2}}}}_j \dots \cstl^{\circled{C_{i_{t}}}}_j $ for $j \in [l]$. Since the output expression $O(q_{s, m_{s}})$ defined above has already taken into consideration the expressions $\cstl^{\circled{H_{s+1}}}_j \dots \cstl^{\circled{H_{r}}}_j$ for $j \in [l]$, in Step II$'$, we can do the analysis as if we have a generalized lasso where the handle is $H_1\dots H_s$, the collection of cycles is $\{C_{s,1},\dots, C_{s,n_s}$, $\dots$, $C_{r,1},\dots, C_{r,n_r}\}$, with the output expression  $O(q_{s, m_{s}})$.
\subsubsection{Step III$'$:}
After Step II$'$, if the decision procedure has not returned yet, then similar to Lemma~\ref{prop-bnd-domain-2}, the following hold.
\begin{itemize}
\item For each $s \in [r]$ and each path $\schm=H_1 \schm_1 H_2 \dots H_s \schm_s$ such that for each $s'\in [s]$, $\schm_{s'}$ is a cycle scheme over the collection of cycles $\{C_{s',1},\dots,C_{s',n_{s'}}\}$, it holds that the constant atom and all the coefficients of the non-constant atoms in ${\sumf^{(\schm,\sval_0)}}^-(y_j)$ are from a bounded domain $U$.
\item Moreover,  an abstraction of $\schm$, denoted by $\abs(\schm)$, can be defined, so that $\mathscr{A}$, which is the set of $\abs(\schm)$ for the paths $\schm=H_1 \schm_1 H_2 \dots H_s \schm_s$ (where $s \in [r]$), can be computed effectively from 
$H_1, C_{1,1}, \dots, C_{1,n_1},H_2,\dots, H_r,C_{r,1},\dots, C_{r,n_r}$.
\end{itemize}
\smallskip
\framebox[\textwidth]{
	\begin{minipage}{0.95\textwidth}
\noindent {\bf Step III$'$}. We apply the same analysis to $\mathscr{A}$ as in Step III. If the procedure does not return during the analysis, then return $\lfalse$.
	\end{minipage}
}\medskip\\

\noindent {\it Complexity analysis of Step I$'$-III$'$}. The complexity of Step I$'$ is polynomial in the the maximum length of generalized multi-lassos in $\Ss$. The complexity of Step II$''$ is exponential in the maximum number of simple cycles in a generalized multi-lasso. The complexity of Step III$'$ is exponential in the number of control and data variables in $\Ss$. In total, the complexity of the decision problem for the non-zero output problem of normalized SNTs is exponential in the number of control and data variables, as well as in the number of simple cycles, in the worst case.

%


\section{Extensions}
\label{sec:cases}

\begin{figure}
	\centering
	\lstset{language=C,
		basicstyle=\ttfamily\scriptsize}
	\begin{tabular}{|c|c|c|}
		\hline
		\begin{minipage}[t]{0.2\textwidth}
		\vspace{-0.5cm}
			\begin{lstlisting}[mathescape=true]
int avg() {
 sum:=$\cur$;
 cnt:=0;$\nnext$;
 loop{
  sum+=$\cur$;
  cnt+=1;
  $\nnext$;}
 ret sum/cnt;}
			\end{lstlisting}
		\end{minipage}&
		\begin{minipage}[t]{0.4\textwidth}
		\vspace{-0.5cm}
\begin{lstlisting}[mathescape=true]
int MAD() {
 sum:=$\cur$;cnt:=0;$\nnext$;
 loop{sum+=$\cur$;cnt+=1;$\nnext$;}
 avg:= sum/cnt;mad:=0;$\init$;
 loop{
  if($\cur$<avg){mad+=avg-$\cur$;}
  else{mad+=$\cur$-avg;}$\nnext$;}
 ret mad/cnt;}
\end{lstlisting}
		\end{minipage}&
		\begin{minipage}[t]{0.4\textwidth}
		\vspace{-0.5cm}
			\begin{lstlisting}[mathescape=true]
int SD() {
 sum:=$\cur$;cnt:=0;$\nnext$;
 loop{sum+=$\cur$;cnt+=1;$\nnext$;}
 avg:= sum/cnt;sd:=0;$\init$;
 loop{
  sd+=($\cur$-avg)*($\cur$-avg);$\nnext$;
 }
 ret SQRT(sd/cnt);}
			\end{lstlisting}
		\end{minipage}\\
		\hline		
	\end{tabular}
	\caption{More Challenging Examples of Reducers Performing Data Analytics Operations}
	\label{fig:examples2}
\vspace{-0.5cm}
\end{figure}

In this section, we discuss some extensions of our approach to deal with the more challenging examples. 
For cases with multiplication, division, or other more complicated functions at the return point, e.g., the \texttt{avg} program, we can model them as an \emph{uninterpreted $k$-ary function} and verify that all $k$ parameters of the uninterpreted functions remain the same no matter how the input is permuted, e.g., the \texttt{avg} program always produces the same \texttt{sum} and \texttt{cnt} for all permutation of the same input data word. This is a \emph{sound} but \emph{incomplete} procedure for verifying programs of this type. Nevertheless, it is not often that a  practical program for data analytics produces, e.g., $2q/2r$ from some input and $q/r$ for its permutation. Hence this procedure is often enough for proving commutativity for real world programs.

The \texttt{MAD} (Mean Absolute Deviation) program is a bit more involved. Beside the division operator $/$ that also occurs in the \texttt{avg} example, it uses a new iterator operation $\init$, which resets $\cur$ to the head of the input data word. The strategy to verify this program is to divide the task into two parts: (1) ensure that the value of \texttt{avg} is independent of the order of the input, (2) treat \texttt{avg} as a control variable whose value is never updated and then check if the 2nd half of the program (c.f., Fig.~\ref{fig:examples3}) is commutative. 

\begin{wrapfigure}{r}{0.4\textwidth}

	\lstset{language=C,
		basicstyle=\ttfamily\scriptsize}
	\begin{lstlisting}[mathescape=true]
int MAD2() {
 avg:= $\cur$;$\nnext$;
 loop{
  if($\cur$<avg){mad+=avg-$\cur$;}
  else{mad+=$\cur$-avg;}
  $\nnext$;}
 ret mad/cnt;}
	\end{lstlisting}	

	\caption{The 2nd half of MAD}
	\label{fig:examples3}

\end{wrapfigure}
We handle the division at the end of the program in Fig.~\ref{fig:examples3} in the same way as we did for the \texttt{avg} program. The guarantee we obtain after the corresponding SNT is checked to be commutative is that the program outputs the same value for any value of \texttt{avg} and any permutation of  the input data word.




The \texttt{SD} (Standard Deviation) program is even more challenging. The main difficulty comes from the use of multiplication in the middle of the program (instead of at the return point). In order to have a sound procedure to verify this kind of programs, we can extend the transitions of SNTs to include uninterpreted $k$-ary functions. However, this is not a trivial extension and we leave it as future work.


\vspace{-0.2cm}
	
\section{Conclusion}
\label{sec:conclusion}


The contribution of the paper is twofold. We propose a verifiable programming language for reducers. Although it is still far away from a practical programming language, we believe that some ideas behind our language (e.g., the separation of control variables and data variables) would be valuable for the design of a practical reducer language. On the other hand, we propose the model of streaming numerical transducers, a transducer model over infinite alphabets. To our best knowledge, this is the first decidable automata model over infinite alphabets that allows linear arithmetics over the input values and the integer variables. Although we required that the transition graphs of SNTs are generalized flat,  SNTs with such kind of transition graphs turn out to be quite powerful, since they are capable of simulating reducer programs without nested loops, which is a typical scenario of reducer programs in practice. At last, we would like to mention that although we assumed the integer data domain, all the results obtained in this paper are still valid when a dense data domain, e.g. the set of rational numbers, is assumed.

\medskip

\noindent {\bf Acknowledgements.} Yu-Fang Chen is partially supported by the MOST project No. 103-2221-E-001-019-MY3. Zhilin Wu is partially supported by the NSFC grants No.\ 61100062, 61272135, 61472474, and 61572478.



\vspace{-2mm}

\bibliographystyle{abbrv}
\bibliography{data}

\newpage


\begin{appendix}

\section{Formal Semantics of the Programming Language}
\begin{figure}
\begin{center}
	\hspace{-0.4cm}
	\scalebox{0.9}{
		\begin{tabular}{|l|l|}
			\hline
			Transitions&
			Side Condition\\
			\hline
			$(y := e;p, w, \rho) \longrightarrow (p, w, \rho')$&
			$\rho'=\rho[\eval{e}{\rho}/y]$\\
			
			
			$(y \addeq e;p, w, \rho) \longrightarrow (p, w, \rho')$&
			$\rho'=\rho[\eval{y+e}{\rho}/y]$\\
			
			$(\ite{g}{s_1}{s_2};p, w, \rho) \longrightarrow (s_1;p, w, \rho)$&
			$\rho \models g$\\
			
			$(\ite{g}{s_1}{s_2};p, w, \rho) \longrightarrow (s_2;p, w, \rho)$& $\rho \not \models g$\\
			
			$(\nnext;p, w, \rho) \longrightarrow (p, \tail(w), \rho')$&
			$\rho'=\rho[\head(w)/\cur]$ if $w \neq \varepsilon$ \\
			
			$(x:=\cur;p, w, \rho) \longrightarrow (p, w, \rho')$&
			$\rho'  =\rho[\rho(\cur)/x]$\\
			
			$(\loopL{s};\mbox{ret }r, w, \rho) \longrightarrow (s;\loopL{s};\mbox{ret }r, w, \rho)$& \\
			

			$(\loopL{s};\mbox{ret }r, \epsilon, \rho) \longrightarrow (\mbox{ret }r,  \epsilon, \rho)$& 	\\	

			\hline
			
		\end{tabular}
	}
	\caption{The Semantics of the Programming Language}
	\label{fig:semantics}
\end{center}
\end{figure}

Formally, the semantics of a program $p$ in the programming language is defined as a transition system in Fig.~\ref{fig:semantics}. Let $p$ be a reducer program and $w$ be an input data word.  Each configuration of the transition system is a triple $(p', w', \rho)$, where $p'$ is a program, $w'$ is a suffix of $w$, and $\rho$ is a valuation over $X^+\cup Y$ such that $\rho(\cur)=\head(w')$ (where if $w'=\varepsilon$, then $\head(w')=\bot$). 
Let $\rho_0$ denote a valuation which assigns each control and data variable an initial value, and $\rho_w$ be the valuation such that $\rho_w(\cur)=\head(w)$ and $\rho_w(z) = \rho_0(z)$ for each $z \in X \cup Y$.
The initial configuration is $(p, \tail(w), \rho_w)$.
We use $p_{\rho_0}(w)$ to denote the \emph{output} of $p$ on $w$ wrt. $\rho_0$. Then $p_{\rho_0}(w) =d$ if there exists a path from the initial configuration $(p, w, \rho_w)$ to some return configuration $(\mbox{ret }r,  \epsilon, \rho_r)$ such that $
\eval{r}{\rho_r}=d$. Otherwise, $p_{\rho_0}(w)=\bot$. Since the program is deterministic, i.e., given an initial valuation $\rho_0$, each input data word has at most one output, the semantics of $p$ is well-defined.

\section{Proofs in Section~\ref{sec:def-snt}}

\newcommand\assume{\mathsf{assume}}

\newcommand\loc{\mathfrak{l}}

\noindent {\bf Proposition~\ref{prop-mrprog-to-snt}}.
{\it 
For each reducer program $p$, one can construct an equivalent SNT $\Ss$ where the number of states and the maximum number of simple cycles in an SCC of the transition graph are at most exponential in the number of branching statements in $p$. 
}

\smallskip

\begin{proof}
We introduce a few notations first.

Let $s$ be a loop-free program. An \emph{execution path} $\pi$ of $s$ is a maximal path in the control flow graph of $s$ (here we use the standard definition of control flow graphs). Each execution path $\pi$ corresponds to a program $s_\pi$ obtained by sequentially composing the statements in $\pi$, where the statements $\assume(g)$ are used to represent the guards $g$. Then $s$ can be seen as a union of $s_\pi$, where $\pi$ ranges over the execution paths of $s$. 

Let $p$ be a reducer program of the form $s_1; \nnext; \loopL{s_2;\nnext;}$; ret $r$.  In the following, we show how to construct an SNT $\Ss_p$ to simulate $p$.

The loop body $s_2;\nnext$ can be seen as a union of programs $p_\pi$ for execution paths $\pi$. We assume that no two distinct programs $p_\pi$ share locations. We first transform the loop into a collection of state-disjoint (except the state $\loc_0$, the entry point of the loop)  cycles $C_\pi$, one for each program $p_\pi$.  Let us focus on a program $p_\pi$. The set of states in $C_\pi$ comprises the location $\loc_0$ which is the entry point of the loop, and the locations succeeding each $\nnext$ statement in $p_\pi$. Moreover, we identify the location succeeding the last $\nnext$ statement and the entry point. The effect of the subprogram $s'$ between two successive $\nnext$ statements in the locations $\loc_1,\loc_2$ can be summarized into a transition  $(\loc_1, g', \eta', \loc_2)$ of $p_\pi$. This is possible due to the following two constraints: 1) the conditions $g$ in the statements $\ite{g}{s'_1}{s'_2}$  of $p_\pi$ are the conjunctions of $\cur \odot c$ and $\cur \odot x$, 2) the assignments to the control variables are of the form $x:=\cur$, and the assignments to the data variables are of the form $y:=e$ and $y {+=} e$, where $e$ contains only control variables or $\cur$. As a result of the two constraints, we can trace the evolvement of the values of the control variables and simulate all the statements $\assume(g)$ occurring in $s'$ by a guard $g'$ (obtained from these guards $g$ by some variable substitutions),  moreover, the effects of all the assignments therein can be summarized into an assignment function $\eta'$.  Similarly, we can do the same for the subprogram between the entry point and the first $\nnext$ statement of $p_\pi$.

In addition, each execution path of $s_1;\nnext;$ can be simulated by a simple path of transitions of $\Ss_p$, which ends in the state $\loc_0$, the entry point of the loop.

The return statement t is handled by adding a transition with guards of the form $\cur=\wend$ from the state $\loc_0$  to a sink state $\loc'$.
The output function $O_p$ of $\Ss_p$ is defined as follow: $O_p(\loc') = r$ and $O(\loc)$ is undefined for all the other states $\loc$.

Because in the program ``$s_1; \nnext; \loopL{s_2;\nnext;}$; ret $r$'', the subprogram $s_2$ contains no occurrences of $\nnext;$, we know that each nontrivial SCC  in $\Ss_p$ comprises a collection of self-loops around a unique state. Therefore, $\Ss_p$ is generalized-flat.
\qed
\end{proof}

\vspace{4mm}

\noindent {\bf Proposition~\ref{prop-snt-cmm-to-eqv}}. 
\emph{The commutativity problem of SNTs is reduced to the equivalence problem of SNTs in polynomial time}.

\begin{proof}
Suppose that $\Ss=(Q, X, Y, \delta, q_0, O)$ is an SNT such that $X=\{x_1,\dots,x_k\}$ and $Y=\{y_1,\dots,y_l\}$. Without loss of generality, we assume that the output of $\Ss$ is defined only for data words of length at least two. We will construct two SNTs $\Ss_1$ and $\Ss_2$ so that $\Ss$ is commutative iff $\Ss$ is equivalent to both $\Ss_1$ and $\Ss_2$.
\begin{itemize}
\item The intuition of $\Ss_1$ is that over a data word $w=d_1 d_2 d_3 \dots d_n\wend$ with $n\ge 2$, $\Ss_1$ simulates the run of $\Ss$ over $d_2 d_1 d_3 \dots d_n\wend$, that is, the data word obtained from $w$ by swapping the first two data values.
\item The intuition of $\Ss_2$ is that over a data word $w=d_1 d_2 d_3 \dots d_n\wend$ with $n\ge 2$, $\Ss_1$ simulates the run of $\Ss$ over $d_2 d_3 \dots d_n d_1\wend$, that is, the data word obtained from $w$ by moving the first data value to the end. 
\end{itemize}
The correctness of this reduction follows from Proposition 1 in \cite{CHSW15}.

\smallskip

\noindent {\it The construction of $\Ss_1$}.

Intuitively, over a data word $w=d_1d_2 d_3 \dots d_n\wend$, we introduce an additional write-once variable $x'$ to store $d_1$, then simulates the run of $\Ss$ over $d_2 d_1 d_3 \dots d_n\wend$ as follows: When reading $d_2$ in $w$, the variables are updated properly by letting $x'$ to represent $d_1$ and $\cur$ to represent $d_2$.

Let $q'_{0},q'_{1} \not \in Q$ and $x' \not \in X$. Then $\Ss_1 = (Q \cup \{q'_{0},q'_1\}, X \cup \{x'\}, Y, \delta_1, q'_{0}, O_1)$ such that 
\begin{itemize}
\item $O_1(q'_0)$ and $O_1(q'_1)$ are undefined, and for each $q \in Q$, $O_1(q)=O(q)$,
\item $\delta_1$ is constructed from $\delta$ as follows,
\begin{itemize}
\item each element of $\delta$ is an element of $\delta_1$,
\item for each pair of transitions $q_0 \xrightarrow{(g_1,\eta_1)} q_1 \xrightarrow{(g_2,\eta_2)} q_2$ in $\Ss$, we add the transitions $(q'_0, \cur\neq \wend, \eta'_1, q'_1)$ and $(q'_1, g'\wedge \cur \neq \wend, \eta'_2, q_2)$ into $\delta_1$. Intuitively, we use $x'$ to store the value of $d_1$ in $ \eta'_1$ and summarize the computation of $\eta_1$ and $\eta_2$ in $\eta'_2$ with the information that $d_1$ is stored in $x'$ and $d_2$ is stored in $\cur$.

Formally, $\eta'_1, g', \eta'_2$ are defined as follows.
\begin{itemize}
\item $\eta'_1(x')=\cur$ and $\eta'_1(z)$ is undefined for all $z \in X\cup Y$. This implies that after the transition $(q'_0, \ltrue, \eta'_1, q'_1)$, each variable $z \in X \cup Y$ still holds the initial value.  
\item $g' = g_1 \wedge g'_2$, where $g'_2$ is obtained from $g_2$ by replacing $\cur$ with $x'$, and each $x \in X\cap \dom(\eta_1)$ with $\eta_1(x)$.
\item For each $z \in X\cup Y$, if $z \in \dom(\eta_2)$, let $\eta_2^r(z)$ be the expression obtained by replacing all occurrences of $\cur$ in $\eta_2(z)$ with $x'$,
then $\eta'_2(z)$ is obtained by substituting all occurrences of variables $z' \in \dom(\eta_1)$ in $\eta_2^r(z)$ with $\eta_1(z')$. 
\item For each $z \in X\cup Y$, if $z \notin \dom(\eta_2)$, then $\eta'_2(z)=\eta_1(z)$.

\hide{
\item For each $y_j \in Y$, if $y_j \in \dom(\eta_2)$, then 
\[
\begin{array}{l c l}
\eta'_2(y_j) & = & (a_{j} + b_{j}\cur) + a'_{j} + b'_{j,0} x' + \sum \limits_{x_{j'} \in \dom(\eta_1)} b'_{j,j'} \cur \\
& = & (a_{j} + a'_{j}) + b'_{j,0} x' + (b_{j}  + \sum \limits_{x_{j'} \in \dom(\eta_1)} b'_{j,j'} )\cur,
\end{array}
\]
(or 
\[
\begin{array}{l c l}
\eta'_2(y_j) & = & a'_{j} + b'_{j,0} x' + \sum \limits_{x_{j'} \in \dom(\eta_1)} b'_{j,j'} \cur  \\
& = & a'_{j} + b_{j,0} x' + (\sum \limits_{x_{j'} \in \dom(\eta_1)} b'_{j,j'}  )\cur.
\end{array}
\], respectively).

Otherwise, if $y_j \in \dom(\eta_1)$, then $\eta'_2(y_j)= a_{j} + b_{j} \cur$. Otherwise, $\eta'_2(y_j)$ is undefined.}
\end{itemize}
\end{itemize}
\end{itemize}

\smallskip

\noindent {\it The construction of $\Ss_2$}.

Intuitively, over a data word $w=d_1\dots d_n\wend$, we introduce an additional control variable $x'$ to store $d_1$, then simulates the run of $\Ss$ over $d_2\dots d_n d_1$: When reading $\wend$ in $w$, the variables are updated properly by letting $x'$ to represent $d_1$.


Suppose $q'_{0} \not \in Q$ and $x' \not \in X$. Then $\Ss_2 = (\{q'_0\} \cup Q, X\cup\{x'\}, Y, \delta_2, q'_{0}, O_2)$ such that  
\begin{itemize}
\item $O_2(q'_0)$ is undefined, and for each $q \in Q$, $O_2(q)=O(q)$,	
	
\item $\delta_2$ is constructed from $\delta$ as follows,
	\begin{itemize}
	\item each element of $\delta$ whose guard does not contain $\cur = \wend$ is an element of $\delta_2$,

	\item we add the transition $q'_0 \xrightarrow{(\ltrue,\eta'_1)} q_0$ to $\delta_2$, where $\eta'_1(x') =\cur$ and $\eta'_1(z)$ is undefined for all $z \in X\cup Y$.

	\item for each pair of transitions $q_1 \xrightarrow{(g_1 \wedge \cur \neq \wend, \eta_1)} q_2 \xrightarrow{(g_2\wedge \cur=\wend, \eta_2)} q_3$ in $\Ss$, we add the transition $(q_1, g', \eta'_2, q_3)$ into $\delta_2$. Intuitively, we use $x'$ to store the value of $d_1$ in $ \eta'_1$ and summarize the computation of $\eta_1$ and $\eta_2$ in $\eta'_2$ with the information that $d_1$ is stored in $x'$.

	Formally, $g',\eta'_2$ are defined in the following. 
		\begin{itemize}
		\item $g' = g'_1 \wedge g'_2 \wedge(\cur=\wend)$, where $g'_1$ and $g'_2$ are obtained from $g_1,g_2$ as follows: $g'_1$ is obtained from $g_1$ by replacing all occurrences of $\cur$ with $x'$, and $g'_2$ is obtained from $g_2$ by replacing each $x \in X\cap \dom(\eta_1)$ with $\eta_1(x)$, then substituting all occurrences of $\cur$ with $x'$.
		\item For each $z \in X\cup Y$, if $z \in \dom(\eta_2)$, then $\eta'_2(z)$ is the expression obtained from $\eta_2(z)$ by replacing all occurrences of variables $z' \in \dom(\eta_1)$ therein with $\eta_1(z')$ and then substituting all occurrences of $\cur$ with $x'$. 
		\item For each $z \in X\cup Y$, if $z \in \dom(\eta_1) \setminus \dom(\eta_2)$, then $\eta'_2(z)$ is the expression obtained from $\eta_1(z)$ by substituting all occurrences of $\cur$ with $x'$.
		\item For each $z \in X\cup Y$, if $z \notin \dom(\eta_2) \cup \dom(\eta_1)$, then $\eta'_2(z)$ is undefined.

		\end{itemize}
	\end{itemize}
\end{itemize}

It is easy to see that the size of both $\Ss_1$ and $\Ss_2$ are polynomial with respect to the size of $\Ss$.
Note that $\Ss_1$ and $\Ss_2$ constructed above preserve the generalized-flatness and mononicity of $\Ss$, since the constructions do not modify the transitions in the nontrivial SCCs of the transition graph.
\qed
\end{proof}

\noindent {\bf Proposition \ref{prop-snt-eqv-to-nzero}}.
\emph{From two SNTs $\Ss_1$ and $\Ss_2$, an SNT $\Ss_3$ can be constructed in polynomial time such that  $(\Ss_1)_{\rho_0}(w\wend) \neq (\Ss_2)_{\rho_0}(w\wend)$ for some  data word $w \wend$ and valuation $\rho_0$  iff $(\Ss_3)_{\rho_0}(w\wend) \not\in \{\bot,0\}$ for some data word $w\wend$ and valuation $\rho_0$.}

\begin{proof}
Let $\Ss_1 = (Q_1,X_1,Y_1,\delta_1,q_{1,0}, O_1)$ and  $\Ss_2 = (Q_2,X_2,Y_2,\delta_2,q_{2,0}, O_2)$ be two monotone SNTs. Without loss of generality, we assume that $Q_1 \cap Q_2 = \emptyset$, $X_1 \cap X_2 = \emptyset$, and $Y_1 \cap Y_2 = \emptyset$. 

We first construct $\Ss$ as the product of $\Ss_1$ and $\Ss_2$. Specifically, $\Ss=(Q_1 \times Q_2, X_1 \cup X_2, Y_1 \cup Y_2, \delta, (q_{1,0},q_{2,0}), O)$, where
\begin{itemize}
\item $\delta$ comprises $((q_1,q_2), g_1 \wedge g_2, \eta_1 \cup \eta_2, (q'_1,q'_2))$ such that $(q_1,g_1,\eta_1,q'_1) \in \delta_1$, $(q_2,g_2,\eta_2,q'_2) \in \delta_2$, and $g_1 \wedge g_2$ is satisfiable,
\item for each $(q_1,q_2) \in Q_1 \times Q_2$, 
\begin{itemize}
\item if $O_1(q_1)$ is defined and $O_2(q_2)$ is undefined or vice versa, then $O((q_1,q_2))=1$, 
\item otherwise, if both $O_1(q_1)$ and $O_2(q_2)$ are defined, then $O((q_1,q_2))=O_1(q_1) - O_2(q_2)$, 
\item otherwise (both $O_1(q_1)$ and $O_2(q_2)$ are undefined), $O((q_1,q_2))$ is undefined. 
\end{itemize}
\end{itemize}
From the aforementioned construction, it is easy to see that $\Ss_1$ and $\Ss_2$ are  inequivalent iff there is a data word $w$ such that the output of $\Ss$ over $w$ is non-zero. Moreover, since both $\Ss_1$ and $\Ss_2$ are generalized-flat and monotone, we know that $\Ss$ is generalized-flat and monotone as well.   \qed
\end{proof}

\noindent {\bf Proposition~\ref{prop-snt-norm}}.
{
For each SNT $\Ss$, the nonzero-output problem of $\Ss$ can be reduced to a series of the nonzero-output problems of normalized SNTs $\Ss'$ in exponential time.
}
\smallskip

\newcommand{\tpo}{\mathsf{TPO}}

\begin{proof}
Suppose $\Ss = (Q, X, Y, \delta, q_0, O)$ is an SNT such that $X=\{x_1,\dots, x_k\}$. Our goal is to reduce the nonzero-output problem of $\Ss$ to a series of  nonzero-output problems of normalized SNTs $\Ss'  = (Q', X, Y, \delta', q'_0, O')$.



Let $\tpo_{X}$ denote the set of total preorders between control variables (Recall that a total preorder over $X$ is reflexive and transitive relation $\preceq$  over $X$ such that for each $x_i, x_j \in X$, either $(x_i, x_j) \in \preceq$ or $(x_j, x_i) \in \preceq$). For $\preceq \in \tpo_X$ and $x_i, x_j \in X$,  $x_i$ is said to be a \emph{$\preceq$-successor} of  $x_j$ or $x_j$ is said to be a \emph{$\preceq$-predecessor} of $x_i$, if $(x_j, x_i) \in \preceq$, $(x_i, x_j) \not \in \preceq$, and for each $x_{i'} \in X$ such that $(x_j, x_{i'}) \in \preceq$, it holds that $(x_i, x_{i'}) \in \preceq$. 

Then for each $\preceq_0 \in \tpo_X$, we construct an SNT $\Ss'= (Q', X, Y, \delta', q'_0, O')$ as follows: $Q'= Q \times \tpo_X$, and $q'_0=(q_0,  \preceq_0)$. Moreover, $O'$ is defined as follows: For each $(q, \preceq) \in Q'$, $O'((q, \preceq)) = O(q)$. It remains to define $\delta'$.

The transition set $\delta'$ is defined by the following rules:
\begin{itemize}
\item For each $(q, g \wedge \cur \neq \triangleright, \eta, q') \in \delta$, $\delta'$ includes all the transitions $(q, \preceq) \xrightarrow{(g' \wedge \cur \neq \triangleright, \eta)} (q', \preceq')$ satisfying the following constraints. 
\begin{itemize}
\item  $g'$ is of one of the following forms, 
\begin{itemize}
\item $g' \equiv \varphi_{\preceq} \wedge \cur = x_i$, or
\item $g' \equiv \varphi_{\preceq} \wedge \cur > x_j \wedge \cur < x_i$ such that $x_i \in X$ is a $\preceq$-successor of $x_j \in X$, or
\item $g' \equiv \varphi_{\preceq} \wedge \cur > x_i$ such that there does not exist a $\preceq$-successor of $x_i$, or
\item $g' \equiv \varphi_{\preceq} \wedge \cur < x_i$ such that there does not exist a $\preceq$-predecessor of $x_i$,
\end{itemize}
where $\varphi_{\preceq} \equiv \bigwedge \limits_{ i' < j'} \psi_{x_{i'}, x_{j'}}$, and for each pair of indices $i', j' \in [k]$ such that $i' < j'$,  if $(x_{i'}, x_{j'}), (x_{j'}, x_{i'}) \in \preceq$, then $\psi_{x_{i'}, x_{j'}} \equiv x_{i'} = x_{j'}$, otherwise, if $(x_{i'}, x_{j'}) \in \preceq$ and $(x_{j'}, x_{i'}) \not \in \preceq$, then $\psi_{x_{i'}, x_{j'}} \equiv x_{i'} < x_{j'}$, otherwise, $\psi_{x_{i'}, x_{j'}} \equiv x_{i'} > x_{j'}$.
\item $g$ and $g'$ are compatible, more precisely, $g \wedge g'$ is satisfiable (note that $\varphi_{\preceq}$, which encodes the information in $\preceq$, has been included in $g'$). 

%
%
%
%
\item $\preceq'$ is constructed as follows.
\begin{itemize}
\item Case $g' \equiv \varphi_{\preceq} \wedge \cur = x_i$: At first, for each $x_{i'} \in X$, introduce a fresh variable $x'_{i'}$ to denote the value of $x_{i'}$ after the transition $(q, g \wedge \cur \neq \triangleright, \eta,q')$. Let $X'$ denote the set of fresh variables. Let $\preceq''$ be the reflexive and transitive closure of the relation
\[
\begin{array}{l}
\preceq \cup\ \{(x_{i'}, x'_{i'}), (x'_{i'}, x_{i'}) \mid x_{i'} \in X \setminus \dom(\eta)\}\ \cup\\
 \{(\cur, x_i), (x_i, \cur)\}\ \cup   \{(x'_{i'}, x_{i''}), (x_{i''}, x'_{i'}) \mid \eta(x_{i'})=x_{i''}\}\  \cup \\
 \{(x'_{i'}, \cur), (\cur, x'_{i'}) \mid \eta(x_{i'}) = \cur \}.
\end{array}
\] 
Then $\preceq'$ is the total preorder obtained from $\preceq'' \ \cap\ X' \times X'$ by replacing each $x'_{i'} \in X'$ with $x_{i'} \in X$.

\item Case $g' \equiv \varphi_{\preceq} \wedge \cur > x_j \wedge \cur < x_i$:
At first, for each $x_{i'} \in X$, introduce a fresh variable $x'_{i'}$ to denote the value of $x_{i'}$ after the transition $(q, g \wedge \cur \neq \triangleright, \eta, q')$. Let $X'$ denote the set of fresh variables. Let $\preceq''$ be the reflexive and transitive closure of the relation
\[
\begin{array}{l}
\preceq \cup\ \{(x_{i'}, x'_{i'}), (x'_{i'}, x_{i'}) \mid x_{i'} \in X \setminus \dom(\eta)\}\ \cup\\
 \{(x_j, \cur), (\cur, x_i)\}\ \cup \{(x'_{i'}, x_{i''}), (x_{i''}, x'_{i'}) \mid \eta(x_{i'})=x_{i''}\}\ \cup\\
  \{(x'_{i'}, \cur), (\cur, x'_{i'}) \mid \eta(x_{i'}) = \cur \}.
\end{array}
\] 
Then $\preceq'$ is the total preorder obtained from $\preceq'' \ \cap\ X' \times X'$ by replacing each $x'_{i'} \in X'$ with $x_{i'} \in X$.
\end{itemize}
\item Case $g' \equiv  \varphi_{\preceq} \wedge  \cur > x_i$, or $g' \equiv \varphi_{\preceq} \wedge  \cur < x_i$: Similar.
%
%
\end{itemize}

\item For each $(q, g \wedge \cur = \triangleright, \eta, q') \in \delta$, $\delta'$ includes all the transitions $(q, \preceq) \xrightarrow{(\cur = \triangleright, \eta)} (q', \preceq')$ satisfying the following constraints.
\begin{itemize}
\item $\preceq$ and $g$ are compatible, more precisely, $\varphi_{\preceq} \wedge g$ is satisfiable, where $\varphi_{\preceq}$ is defined as above.
%
\item $\preceq'$ is constructed as follows.
At first, for each $x_{i'} \in X$, introduce a fresh variable $x'_{i'}$ to denote the value of $x_{i'}$ after the transition $(q, g \wedge \cur = \triangleright, \eta, q)$. Let $X'$ denote the set of fresh variables. Let $\preceq''$ be the reflexive and transitive closure of the relation
$
\preceq \cup\ \{(x_{i'}, x'_{i'}), (x'_{i'}, x_{i'}) \mid x_{i'} \in X \setminus \dom(\eta)\}  \cup 
 \{(x'_{i'}, x_{i''}), (x_{i''}, x'_{i'}) \mid \eta(x_{i'})=x_{i''}\}.
$ 
Then $\preceq'$ is the total preorder obtained from $\preceq'' \ \cap\ X' \times X'$ by replacing each $x'_{i'} \in X'$ with $x_{i'} \in X$.
\end{itemize}
\end{itemize}

At first, from the construction, we know that $\Ss'$ is path-feasible, state-dominating, and $\triangleright$-transition-guard-tree. We then show that $\Ss'$ is generalized-flat. It is sufficient to prove that for each state $q$ in some nontrivial SCC $S$ of $\Ss$, there does not exist a nontrivial SCC in $\Ss'$ that includes at least two distinct states $(q, \preceq_1)$ and $(q, \preceq_2)$.

To the contrary, suppose that there are a state $q$ in some nontrivial SCC $S$ of $\Ss$ and two distinct states $(q, \preceq_1)$ and $(q, \preceq_2)$ in some nontrivial SCC of $\Ss'$.

Since $\preceq_1 \neq \preceq_2$, without loss of generality, we assume that there are a pair of distinct control variables $x_i, x_j$ such that $(x_i, x_j) \in \preceq_1$ and $(x_i, x_j) \not \in \preceq_2$. 
We introduce some notations first. For $x \in \{x_i, x_j\}$, we say that $x$ computes the minimum (resp. maximum) value in $S$ if whenever $\cur < x$ (resp. $\cur > x$) occurs in a transition $(q, g, \eta, q)$ of $S$, it holds that $\eta(x)=\cur$. 
We distinguish between the following situations.
\begin{itemize}
\item Suppose that both $x_i$ and $x_j$ compute the minimum value in $S$. Since both $x_i$ and $x_j$ compute the minimum value in $S$, when starting from some configuration $(q, \rho)$ such that $\rho \models x_i \le x_j$ and keep applying the transitions in $S$, we know that in each transition, 
\begin{itemize}
\item either the current data value is less or equal to both $x_i$ and $x_j$, then both $x_i$ and $x_j$ are assigned to the current data value and become equal, 
\item or the current data value is greater than $x_i$ and less or equal to $x_j$, then the current data value is assigned to $x_j$ (with the value of $x_i$ unchanged), then $x_i < x_j$ holds after the transition,
\item or the current data value is greater than both $x_i$ and $x_j$, then both the value of $x_i$ and that of $x_j$ are unchanged.
\end{itemize}
Therefore, when following a path from $(q, \preceq_1)$ to $(q, \preceq_2)$ in $\Ss'$, the fact $x_i \le x_j$ persists. This implies that $(x_i, x_j) \in \preceq_2$, a contradiction.
\item Suppose that both $x_i$ and $x_j$ compute the maximum value in $S$. Similarly to the arguments in the previous situation, we know that when following a path from $(q, \preceq_1)$ to $(q, \preceq_2)$ in $\Ss'$, the fact $x_i \le x_j$ persists. This implies that $(x_i, x_j) \in \preceq_2$, a contradiction.
\item Suppose that $x_i$ computes the minimum value in $S$ and $x_j$ computes the maximum value in $S$. Since $x_i$ computes the minimum value and $x_j$ computes the maximum value in $S$, we know that the value of $x_i$ is non-increasing and the value of $x_j$ is non-decreasing. Therefore, when following a path from $(q, \preceq_1)$ to $(q, \preceq_2)$, the fact $x_i \le x_j$ persists. This implies that $(x_i, x_j) \in \preceq_2$, a contradiction.
\item Suppose that $x_i$ computes the minimum value in $S$ and $x_j$ computes neither the minimum value nor the maximum value in $S$. Then the value of $x_i$ is non-increasing and the value of $x_j$ is unchanged when staying in $S$. The arguments are similar to the previous case.
\item Suppose that $x_i$ computes neither the minimum value nor the maximum value and $x_j$ computes the maximum value in $S$. Then the value of $x_i$ is unchanged and the value of $x_j$ is non-decreasing when staying in $S$. The arguments are similar to the previous case.
\item Suppose that $x_i$ computes the maximum value in $S$ and $x_j$ computes the minimum value in $S$.  Then in $S$, the value of $x_i$ is non-decreasing and the value of $x_j$ is non-increasing.  From $(x_i, x_j) \in \preceq_1$ and $(x_i, x_j) \not \in \preceq_2$, we know that when following a path from $(q, \preceq_1)$ to $(q, \preceq_2)$ in $\Ss'$, sometime the value of $x_i$ becomes strictly greater than that of $x_j$, and this fact persists afterwards. Therefore, we have $(x_j, x_i) \in \preceq_2$ and $(x_i, x_j) \not \in \preceq_2$. Since in $S$, the value of $x_i$ is non-decreasing and the value of $x_j$ is non-increasing, we know that when following a path from $(q, \preceq_2)$ to $(q, \preceq_1)$ in $\Ss'$, $x_j < x_i$ persists. Therefore, $(x_j, x_i) \in \preceq_1$ and $(x_i, x_j) \not \in \preceq_1$, a contradiction.
\item Suppose that $x_i$ computes the maximum value and  $x_j$ computes neither the minimum value nor the maximum value in $S$. Then the value of $x_i$ is non-decreasing and the value of $x_j$ is unchanged when staying in $S$. The arguments are similar to the previous case.
\item Suppose that $x_i$ computes neither the minimum value nor the maximum value in $S$ and $x_j$ computes the minimum value in $S$. Then the value of $x_i$ is unchanged and the value of $x_j$ is non-increasing when staying in $S$. The arguments are similar to the previous case.
\item Suppose $x_i$ (resp. $x_j$) computes neither  the minimum value nor the maximum value in $S$. Then the value of $x_i$ and $x_j$ are unchanged when staying in $S$. Therefore, if $x_i \le x_j$ holds in the state $(q, \preceq_1)$, then it holds in each state belonging to the same SCC as $(q, \preceq_1)$ in $\Ss'$. In particular, $(x_i, x_j) \in \preceq_2$, a contradiction.
\end{itemize}
Consequently, in each of the situations aforementioned, we always get a contradiction. 
We conclude that the assumption is false and $\Ss'$ is indeed generalized-flat.
\qed
\end{proof}

\section{Proofs in Section~\ref{sec-sum}}

\noindent {\bf Proposition~\ref{prop-sum-path}}.
{
\it Suppose that $P$ is a path starting form $p_0$ and the initial values of $X \cup Y$ are represented by a symbolic valuation $\initval$ such that for each pair of variables $x_i, x_j \in X$, $\sval(x_j)=\sval(x_j)$ iff $x_i \sim_{p_0} x_j$. Then the values of $X \cup Y$ after traversing the path $P$ are specified by a symbolic valuation $\sumf^{(P,\initval)}$ satisfying the following conditions.
\begin{itemize}
\item The set of indices of $X$, i.e., $[k]$, is partitioned into $I^{\circled{P}}_{pe}$ and $I^{\circled{P}}_{tr}$, the indices of \emph{persistent} and \emph{transient} control variables, respectively. A control variable is persistent if it stores the initial value of some control variable after traversing $P$, otherwise, it is transient.
\item For each $x_j\in X$ such that $j \in I^{\circled{P}}_{pe}$, $\sumf^{(P,\initval)}(x_j)=\sval(x_{\pi^{\circled{p_0}}(\pi^{\circled{P}}_{pe}(j))})$, where $\pi^{\circled{P}}_{pe}: I^{\circled{P}}_{pe} \rightarrow [s^{\circled{p_0}}]$ is a mapping from the index of a persistent control variable $x_j$ to the index of the equivalence class such that the initial value of control variables corresponding to this equivalence class is assigned to $x_j$ after traversing $P$.
\item  For each $x_j\in X$ such that $j\in I^{\circled{P}}_{tr}$,
$\sumf^{(P,\initval)}(x_j)=\vard^{\circled{P}}_{\pi^{\circled{P}}_{tr}(j)}$, where $\pi^{\circled{P}}_{tr}: I^{\circled{P}}_{tr} \rightarrow [r^{\circled{P}}]$ is a mapping from the index of a transient control variable to the index of the data value assigned to it.
\item For each $y_j \in Y$, 
\[
 \sumf^{(P,\initval)}(y_j)  =
 \cste^{\circled{P}}_{j} + 
 \cstl^{\circled{P}}_j \initval(y_j)  + 
  \sum\limits_{j'\in [s^{\circled{p_0}}]} \csta^{\circled{P}}_{j,j'}\initval(x_{\pi^{\circled{p_0}}(j')}) +
  \sum\limits_{j''\in [r^{\circled{P}}]}\cstb^{\circled{P}}_{j,j''} \vard^{\circled{P}}_{j''},
\]  
where $\cste^{\circled{P}}_j,\cstl^{\circled{P}}_j, \csta^{\circled{P}}_{j,1},\dots,\csta^{\circled{P}}_{j, s^{\circled{p_0}}}, \cstb^{\circled{P}}_{j,1},\dots,\cstb^{\circled{P}}_{j,r^{\circled{P}}}$ are integer constants such that $\cstl^{\circled{P}}_{j} \in \{0,1\}$ (as a result of the ``independently evolving and copyless'' constraint).  It can happen that $\cstl^{\circled{P}}_j =0$,  which means that $\initval(y_j)$ is irrelevant to $\sumf^{(P,\initval)}(y_j)$. Similarly for $\csta^{\circled{P}}_{j,1}=0$, and so on.
\end{itemize}
}

\begin{proof}
Suppose $\Ss=(Q, X, Y, \delta, q_0, O)$ is a normalized SNT and $P=p_0 \xrightarrow{(g_1,\eta_1)} p_1 \dots p_{n-1} \xrightarrow{(g_n,\eta_n)} p_{n}$ is a path of $\Ss$. We assume that the initial values of the control and data variables are represented by a symbolic valuation $\sval$ over $X \cup Y$ such that for each pair of variables $x_i, x_j \in X$, $\sval(x_j)=\sval(x_j)$ iff $x_i \sim_{p_0} x_j$. 

We show by an induction that for each $i: 1 \le i \le n$, a symbolic valuation $\sumf_i$ over $X^+ \cup Y$ can be constructed  to describe the value of $x_j$ (resp. $y_j$) after going through the first $i$ transitions of $P$. Moreover, an index set $I_i \subseteq [k]$ is computed as well. 
%
\begin{itemize}
\item At first, compute $\sumf_0$ and $I_0$ as follows. 
\begin{enumerate}
\item For each $x_j \in X$, $\sumf_0(x_j):=\initval(x_{j'_0})$, where $j'_0= \min(\{j' \in [k] \mid j' \sim_{p_0} j\})$.

\item  If $g_1 \models \cur = x_j$ for some $x_j \in X$, then $\sumf_0(\cur):=\sumf_0(x_{j})$ and $s:=0$, otherwise, $\sumf_0(\cur):=\vard^{\circled{P}}_1$ and $s := 1$.

\item For each $y_j \in Y$, $\sumf_0(y_j):=\initval(y_{j})$.

\item In addition, let $I_0 = \emptyset$.
\end{enumerate}
%
%
\item Let $i: 1 \le i \le n$.  
 Then $\sumf_i$ and $I_i$ are computed as follows: 
\begin{enumerate}
\item Initially, let $I_i := \emptyset$. 

\item For each $x_j \in X$, we distinguish among the following situations,
\begin{itemize}
\item if $x_j \not \in \dom(\eta_i)$, then $\sumf_i(x_j) := \sumf_{i-1}(x_j)$, in addition, if $j \in I_{i-1}$, let $I_i:=I_i \cup \{j\}$, 

\item  if $x_j \in \dom(\eta_i)$, in addition, either $\eta_i(x_j) = x_{j'}$ for some $x_{j'} \in X$, or $\eta_i(x_j) = \cur$ and $\varphi_{q_{i-1}} \wedge g_i \models \cur = x_{j'}$ for some $x_{j'} \in X$, then let $\sumf_i(x_j) := \sumf_{i-1}(x_{j'})$, in addition, if $x_{j'} \in I_{i-1}$,  then let $I_i := I_i \cup \{j\}$,

\item if $\eta_i(x_j) = \cur$ and there do not exist $x_{j'} \in X$ such that $\varphi_{q_{i-1}} \wedge g_i \models \cur = x_{j'}$, then let $\sumf_i(x_j):=\sumf_{i-1}(\cur)$ and $I_i := I_i \cup \{j\}$. 
\end{itemize}
\item Compute $\sumf_i(\cur)$ as follows:
\begin{itemize}
\item If $i< n$ and there exists $x_j \in X$ such that $\varphi_{q_i} \wedge g_{i+1} \models \cur = x_j$, then let $\sumf_i(\cur):=\sumf_{i}(x_{j})$.
\item If $i < n$ and there do not exist $x_j \in X$ such that $\varphi_{q_i} \wedge g_{i+1} \models \cur = x_j$, then let $s:=s+1$ and $\sumf_i(\cur) :=\vard^{\circled{P}}_s$. 

\item If $i = n$, then let $\sumf_i(\cur) := \bot$.
\end{itemize}

\item For each $y_j \in Y$, if $y_j \in \dom(\eta_i)$, then let $\sumf_i(y_j) := \eval{\eta_i(y_j)}{\sumf_{i-1}}$, otherwise, let $\sumf_i(y_j) :=\sumf_{i-1}(y_j)$.
\end{enumerate}
%
%
\end{itemize} 
Then let $I^{\circled{P}}_{tr}:=I_n$, $I^{\circled{P}}_{pe}:=[k] \setminus I^{\circled{P}}_{tr}$, and $r^{\circled{P}}:=s$. The mapping $\pi^{\circled{P}}_{pe}$ and $\pi^{\circled{P}}_{tr}$ are defined as follows: 
\begin{itemize}
\item For each $j \in I^{\circled{P}}_{pe}$, let $x_{j'} \in X$ such that $\sumf_n(x_j) = \initval(x_{j'})$, then $\pi^{\circled{P}}_{pe}(j) := (\pi^{\circled{p_0}})^{-1}(j')$. 

\item For each $j \in I^{\circled{P}}_{tr}$, let $s' \in [r^{\circled{P}}]$ such that $\sumf_n(x_j)=\vard^{\circled{P}}_{s'}$, let $\pi^{\circled{P}}_{tr}(j):=s'$. 
\end{itemize}
The symbolic valuation $\sumf^{(P,\initval)}$ can be defined as the restriction of $\sumf_n$ to $X \cup Y$. Since for each assignment $\eta_i$ and $y_j \in Y$, $\eta_i(y_j) = e$ or $\eta_i(y_j) = y_j +e$ for $e \in \Ee_{X^+}$, it follows that $\sumf^{(P,\initval)}(y_j)$ is of the form required by the proposition.
\qed
\end{proof}

\noindent {\bf Proposition~\ref{prop-sum-cycle}}.
{\it 
Suppose that $C$ is a simple cycle (i.e. a self-loop around a state $q$) and $P=C^{\ell}$ such that $\ell \ge 2$. Then the symbolic valuation $\sumf^{(C^\ell,\initval)}$ to summarize the computation of $\Ss$ on $P$ is as follows:  

\noindent
\medskip
\resizebox{0.95\hsize}{!}{
$
\begin{array}{l c l}
\sumf^{(C^\ell,\initval)}(y_j)  &= & 
\left(1 + \cstl^{\circled{C}}_{j} + \dots +(\cstl^{\circled{C}}_{j})^{\ell - 1} \right)\cste^{\circled{C}}_{j} + (\cstl^{\circled{C}}_{j})^\ell \initval(y_j)\ + \\
\smallskip
&& \sum \limits_{j' \in \rng(\pi^{\circled{C}}_{pe}) } \left(1+\cstl^{\circled{C}}_{j} + \dots +(\cstl^{\circled{C}}_{j})^{\ell - 1} \right) \csta^{\circled{C}}_{j,j'}\initval(x_{\pi^{\circled{q}}(j')}) \ + \\
& &  \sum \limits_{j' \in [s^{\circled{q}}] \setminus \rng(\pi^{\circled{C}}_{pe}) }  (\cstl^{\circled{C}}_{j})^{\ell - 1} \csta^{\circled{C}}_{j,j'} \initval(x_{\pi^{\circled{q}}(j')})\ +  \\
\smallskip
&&  \sum \limits_{j' \in \rng(\pi^{\circled{C}}_{tr})} \sum \limits_{s\in[\ell -1]}
\left(\cstl^{\circled{C}}_{j}\cstb^{\circled{C}}_{j,j'}+ \sum \limits_{j'' \in (\pi^{\circled{C}}_{tr})^{-1}(j') \cap \rng(\pi^{\circled{q}})} \csta^{\circled{C}}_{j, (\pi^{\circled{q}})^{-1}(j'')}  \right)
(\cstl^{\circled{C}}_{j})^{\ell-s-1}
\vard^{\circled{C , s}}_{j'}\ +\\
\smallskip
&& \sum \limits_{j' \in [r^{\circled{C}}] \setminus \rng(\pi^{\circled{C}}_{tr})}\sum \limits_{s\in[\ell -1]} \left((\cstl^{\circled{C}}_{j})^{\ell - s} \cstb^{\circled{C}}_{j,j'} \right) \vard^{\circled{C , s}}_{j'} + 
\sum \limits_{j' \in [r^{\circled{C}}] }  
 \cstb^{\circled{C}}_{j, j'} \vard^{\circled{C , \ell}}_{j'},
\end{array} 
$
}
\medskip\\
where the variables $\vard^{\circled{C , s}}_{1}$ for $s\in [\ell]$
 represent the data values introduced when traversing $C$ for the $s$-th time. 
}

\begin{proof}
We prove by an induction on $\ell$ that $\sumf^{(C^\ell,\initval)}(y_j)$ is of the desired form required by the proposition.

\noindent The induction base: $\ell=2$.

\smallskip

Let $\vard^{(\circled{C, 2})}_{1}, \dots, \vard^{(\circled{C, 2})}_{r^{\circled{C}}}$ be the data values introduced when traversing the cycle for the second time. Then from Corollary~\ref{cor-comp-two-paths}, we know that $\sumf^{(C^{2},\initval)} = \sumf^{(C,\sumf^{(C,\initval)})}$ is defined as follows: For each $y_j \in Y$,

\[
\begin{array}{rl}
	\medskip
	\sumf^{(C^{2},\initval)}(y_j) = & 
	\left(\cste^{\circled{C}}_{j}+
	\cstl^{\circled{C}}_{j} \cste^{\circled{C}}_{j}\right)+ \left(\cstl^{\circled{C}}_{j}\right)^2 \initval(y_j)+ \sum \limits_{j' \in \rng(\pi^{\circled{C}}_{pe}) } \left(1+\cstl^{\circled{C}}_{j} \right) \csta^{\circled{C}}_{j,j'}\initval(x_{\pi^{\circled{q}}(j')})\\
	\medskip
	& 
	 + \sum \limits_{j' \in [s^{\circled{q}}] \setminus \rng(\pi^{\circled{C}}_{pe}) } \cstl^{\circled{C}}_{j}  \csta^{\circled{C}}_{j,j'} \initval(x_{\pi^{\circled{q}}(j')}) \\
	&
	+  \sum \limits_{j' \in \rng(\pi^{\circled{C}}_{tr})} 
	\left(\cstl^{\circled{C}}_{j}\cstb^{\circled{C}}_{j,j'}+ \sum \limits_{j'' \in (\pi^{\circled{C}}_{tr})^{-1}(j') \cap \rng(\pi^{\circled{q}})} \csta^{\circled{C}}_{j, (\pi^{\circled{q}})^{-1}(j'')}  \right) \vard^{\circled{C , 1}}_{j'} 
	 \\
	\smallskip
	& 
	+ \sum \limits_{j' \in [r^{\circled{C}}]\setminus \rng(\pi^{\circled{C}}_{tr})} \left( \cstl^{\circled{C}}_{j} \cstb^{\circled{C}}_{j,j'} \right) \vard^{\circled{C,1}}_{j'} +
	
	\sum \limits_{j'\in[r^{\circled{C}}]} \cstb^{\circled{C}}_{j,j'} \vard^{\circled{C,2}}_{j'}.
\end{array}
\]

\noindent Induction step: Let $\ell \ge 3$.

From the induction hypothesis, we know that for each $y_j \in Y$, $\sumf^{(C^{\ell-1},\initval)}(y_j)$ is of the desired form.

From Corollary~\ref{cor-comp-two-paths}, $\sumf^{(C^\ell,\initval)} = \sumf^{(C, \sumf^{(C^{\ell-1},\initval)})}$. Then for each $y_j \in Y$, by unfolding the expressions $\sumf^{(C^{\ell-1},\initval)}(x_{j'})$ for $j' \in [k]$ and $\sumf^{(C^{\ell-1},\initval)}(y_{j''})$ for $j'' \in [l]$ in $\sumf^{(C, \sumf^{(C^{\ell-1},\initval)})}(y_j)$, we can observe that $\sumf^{(C^\ell,\initval)}(y_j)$ is of the desired form.
\qed
\end{proof}

\section{Proofs in Section~\ref{sec-glasso}}

\noindent {\bf Lemma~\ref{prop-cycle-schm}}.
{\it 
Suppose $\schm=C_{i_1}^{\ell_1} C_{i_2}^{\ell_2} \dots C_{i_t}^{\ell_t}$ is a cycle scheme, and $\initval$ is a symbolic valuation representing the initial values of the control and data variables such that for each $x_i, x_j \in X$, $\initval(x_i) = \initval(x_j)$ iff $i \sim_{q_m} j$. 
For all $j' \in  I^{\circled{C_{i_{1}}}}_{pe} \cap \rng(\pi^{\circled{q_m}})$, let $r_{j'}$ be the largest number $r \in [t]$ such that $j'\in\bigcap_{s\in[r]} I^{\circled{C_{i_{s}}}}_{pe}$, i.e., $x_{j'}$ remains persistent when traversing $C_{i_1}^{\ell_1} C_{i_2}^{\ell_2} \dots C_{i_{r_{j'}}}^{\ell_{r_{j'}}}$.
Then for each $j\in [l]$ and $j' \in I^{\circled{C_{i_1}}}_{pe}  \cap \rng(\pi^{\circled{q_m}})$, the coefficient of the $\initval(x_{j'})$-atom in $\sumf^{(\schm,\initval)}(y_j)$ is 
\begin{center}
\resizebox{0.8\hsize}{!}{
$e+\sum\limits_{s_1\in[r_{j'}]}  
\left(1+\lambda^{\circled{C_{i_{s_1}}}}_{j} + \dots + (\lambda^{\circled{C_{i_{s_1}}}}_{j})^{\ell_{s_1}-1} \right) \csta^{\circled{C_{i_{s_1}}}}_{j,(\pi^{\circled{q_m}})^{-1}(j')}\prod\limits_{{s_2}\in[{s_1}+1,t]}\left(\lambda^{\circled{C_{i_{s_2}}}}_{j}\right)^{\ell_{s_2}}$},
\end{center}
where (1) $e\!=\!0$ when $r_{j'}\!=\!t$ and (2) $e=(\lambda^{\circled{C_{i_s}}}_{j})^{\ell_s-1} \csta^{\circled{C_{i_{s}}}}_{j, (\pi^{\circled{q_m}})^{-1}(j')} \prod\limits_{{s'}\in[s+1,t]}\left(\lambda^{\circled{C_{i_{s'}}}}_{j}\right)^{\ell_{s'}}$ with $s=r_{j'}+1$ when $r_{j'}<t$.\\
The constant atom of $\sumf^{(\schm,\initval)}(y_j)$ is 
\begin{center}
\resizebox{0.7\hsize}{!}{$
\sum\limits_{{s_1}\in[t]}
\left(1+\lambda^{\circled{C_{i_{s_1}}}}_{j} + \dots + (\lambda^{\circled{C_{i_{s_1}}}}_{j})^{\ell_{s_1}-1} \right)
\cste^{\circled{C_{i_{s_1}}}}_{j} 
\prod\limits_{{s_2}\in[{s_1}+1,t]}\left(\lambda^{\circled{C_{i_{s_2}}}}_{j}\right)^{\ell_{s_2}}$}
\end{center}
Moreover, for all $j\!\in\! [l]$, in $\sumf^{(\schm,\initval)}(y_j)$, only the constant atom and the coefficients of the $\initval(x_{j'})$-atoms with $j' \!\in\! I^{\circled{C_{i_1}}}_{pe} \cap \rng(\pi^{\circled{q_m}})$ contain a subexpression of the form $ \mu_\schm \ell_1$ for some~$\mu_\schm\in \intnum$.
}

\begin{proof}
The lemma can be shown by applying Proposition~\ref{prop-sum-cycle}, Corollary~\ref{cor-comp-two-paths}, and an induction on the length $t$ of the cycle schemes.\qed
\end{proof}

\end{appendix}



\end{document}